\documentclass[%
 reprint,
superscriptaddress,
 amsmath,amssymb,
 aps,
pra,
floatfix,
]{revtex4-2}

\usepackage{graphicx}
\usepackage{dcolumn}
\usepackage{bm}

\usepackage{amsmath}
\usepackage{xcolor}

\newcommand{\be} {\begin{equation}}
\newcommand{\ee} {\end{equation}}
\newcommand{\ben} {\begin{equation*}}
\newcommand{\een} {\end{equation*}}
\newcommand{\eq}[1]{\begin{align}#1\end{align}}

\newcommand{\ba} {\begin{array}}
\newcommand{\ea} {\end{array}}
\newcommand{\citeasnoun}[1]{Ref.~\citenum{#1}}

\newcommand{\figref}[1]{Fig.~\ref{#1}}
\newcommand{\Figref}[1]{Figure~\ref{#1}}

\newcommand{\eqrefs}[2]{(\ref{#1}--\ref{#2})}

\newcommand{\rev}[1]{\textcolor{black}{#1}}

\begin{document}


\title{\rev{Maximizing} performance of quantum cascade laser-pumped molecular lasers}

\author{Fan Wang}
\affiliation{Department of Mathematics, Massachusetts Institute of Technology, Cambridge, MA 02139 USA}
\author{Steven G. Johnson}%
\affiliation{Department of Mathematics, Massachusetts Institute of Technology, Cambridge, MA 02139 USA}
\affiliation{Department of Physics, Massachusetts Institute of Technology, Cambridge, MA 02139 USA}
\author{Henry O. Everitt}%
\affiliation{Department of Physics, Duke University, Durham, NC 27708, USA}
\affiliation{U.S. Army CCDC Aviation \& Missile Center, Redstone Arsenal, AL 35898, USA}

\date{\today}

\begin{abstract}
Quantum cascade laser \rev{(QCL)}-pumped molecular lasers (QPMLs) have recently been introduced as a new source of powerful ($>$ 1 mW), tunable ($>$ 1 THz), narrow-band ($<$ 10 kHz), continuous-wave terahertz radiation.  The performance of these lasers depends critically on molecular collision physics, pump saturation, and on the design of the laser cavity.  Using a validated three-level model that captures the essential collision and saturation behaviors of the QPML gas nitrous oxide (N$_2$O), we explore how threshold pump power and output terahertz power depend on pump power, gas pressure, as well as on the diameter, length, and output-coupler transmissivity of a cylindrical cavity.  The analysis indicates that maximum power occurs as pump saturation is minimized in a manner that depends much more sensitively on pressure than on cell diameter, length, or transmissivity. \rev{A near-optimal compact laser cavity can produce more than 10~mW of power tunable over frequencies above 1~THz when pumped by a multi-watt QCL.} 

\end{abstract}

\maketitle


\section{Introduction}

The need for powerful, tunable, narrow-band sources to span the ``terahertz gap'' between 0.3--3.0~THz continues to grow as next generation wireless-communication systems demand increasing bandwidth and new opportunities arise for intrinsically short-range communications facilitated by strong, frequency-dependent atmospheric water vapor absorption~\cite{nagatsuma2016advances}.  Most terahertz sources fall into one of two categories: lower-frequency electronic sources such as microwave oscillators or frequency multipliers~\cite{maestrini2010frequency}, backward-wave oscillators~\cite{ives2003development}, and gyrotrons\cite{booske2011vacuum} whose power decreases with increasing frequency; and higher-frequency optical sources such as terahertz quantum cascade lasers~\cite{kohler2002terahertz} and difference-frequency lasers~\cite{mcintosh1995terahertz,evenson1984tunable,inguscio1994isotopic} whose power increases with increasing frequency.  The region in between, where powerful sources are lacking, is known as the terahertz gap.

Recently, one of the original sources of terahertz radiation, gas-phase optically-pumped far-infrared lasers (OPFIRLs), has experienced a revival as the line-tunable CO$_2$ pump laser has been replaced by a continuously tunable quantum cascade laser (QCL)~\cite{pagies2016continuous,pagies2016low,pagies2017progress,Science2019}. By this substitution, virtually any rotational transition of a gas-phase molecular gain medium may be made to lase if pumped by a QCL tuned into coincidence with the associated infrared molecular rotation--vibration transition.  Unlike electronic sources, the Manley--Rowe effect in these QCL-pumped molecular lasers (QPMLs) causes the power to increase with increasing frequency. Consequently, reports of power exceeding 1~mW~\cite{lampin2020quantum} and tunability approaching 1~THz~\cite{Science2019} from QPMLs foreshadow the potential of these sources to span the terahertz gap.

In order to ascertain the potential of these sources, we must explore how the output power may be maximized.  It is well known that the threshold pump power and terahertz output power depend sensitively on the pressure of the molecular gain medium~\cite{chang1970cw,Everitt1986,McCormick1987,Everitt1995,Chua2011,Wang2018,Science2019}. What is not as well known is how output power depends on cavity geometry (diameter, length, output-coupler transmissivity) and how those parameters determine the optimal pressure for obtaining maximal power.  Recently, a comprehensive, physics-based, experimentally-validated model was developed~\cite{Wang2018} that fully captures the interplay of thousands of thermally populated rotational states connected by millions of collisional energy transfer channels in the laser. However, the complexity of that model limits its utility for building the intuition and analytical insights necessary to understand the sensitivity of output power on these factors. Alternatively, in Ref.~\cite{Science2019} a simple three-level model was proposed to describe the essential lasing behavior for the compact QPMLs in the limiting case of very low pressures for which the dominant relaxation mechanism was molecule--wall collisions and $\alpha_\mathrm{IR} L \ll 1$, where $\alpha_\mathrm{IR}$ is the IR pump absorption coefficient and $L$ is the cavity length. 

To capture the behavior of QPMLs at higher pressures where maximal terahertz power may be obtained, a more rigorous version of this three-level model, which includes  dipole--dipole collisions, molecule--wall collisions, and pump saturation, is presented here. This new treatment, validated against the comprehensive model, allows us to study the laser performance across the entire pressure regime and to provide deep insights for optimizing cavity design.  \rev{After using this improved model to identify optimal cavity geometry, we show that a much smaller cavity the size of a cigar can achieve 95\% of the optimal performance, a necessary step toward the technological viability of the QPML concept. We then demonstrate the potential for QPMLs to produce 10's of milliwatts of continuous wave, narrow band power tunable over frequencies above 1 THz using multi-watt QCLs that are or soon will become commercially available.}

\begin{figure}[htp]
\centering
\includegraphics[width=0.5\textwidth]{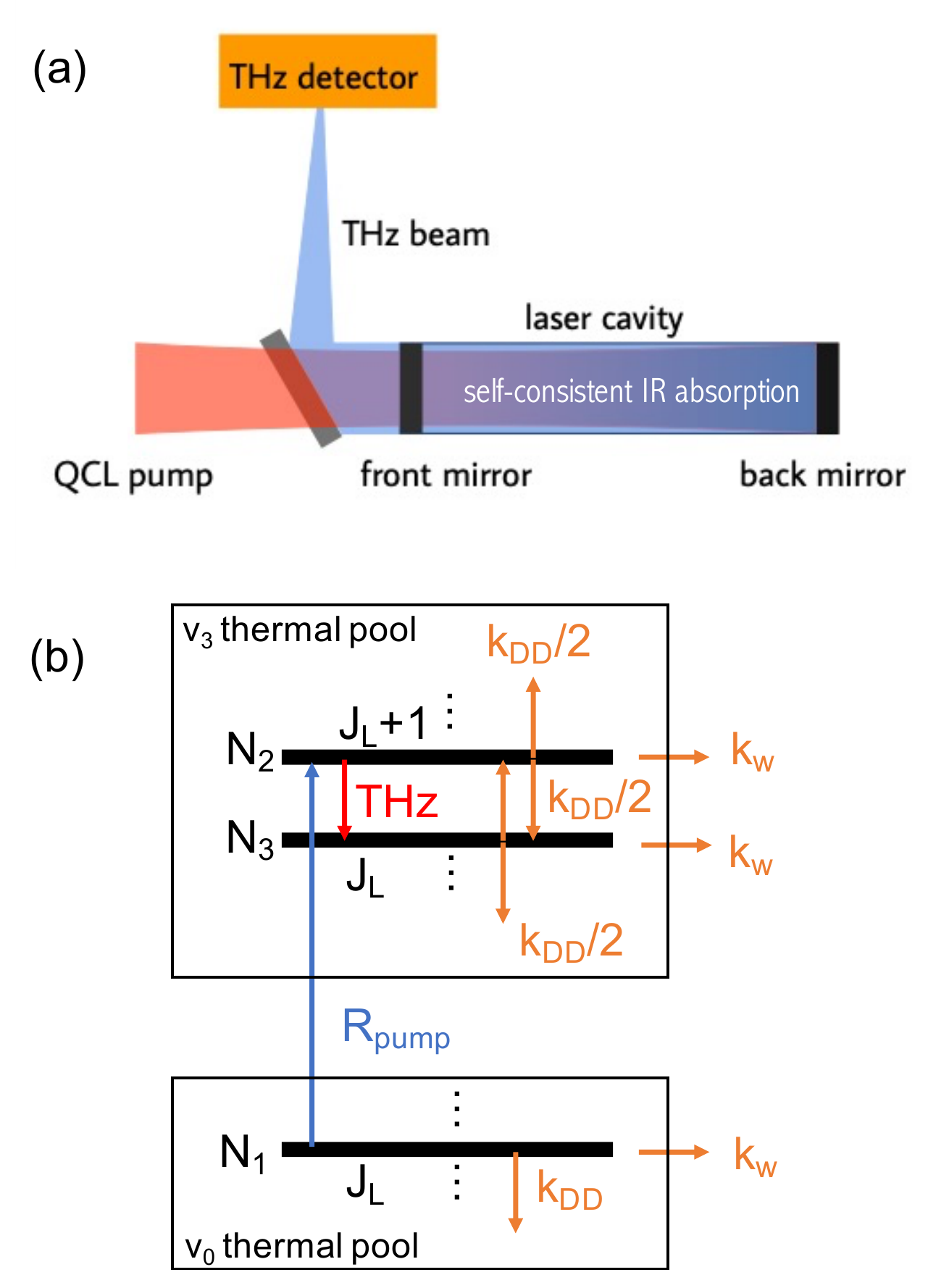}
\caption{(a) Schematics of the QCL-pumped molecular laser. The QCL pumps a cylindrical metallic laser cavity with radius $R_\mathrm{cell}$ and length $L$ filled with molecular gas with pressure $p$. If the pump power exceeds the threshold, terahertz radiation is generated and emitted through a front mirror with transmission coefficient $T$.
(b) A  three-level model that describes the essential physics of the QPML system. 
Molecules in rotational level~1 from the ground vibrational state v$_0$ are pumped into rotational level~2 of the excited vibrational state v$_3$ by the infrared QCL laser, generating terahertz emission between levels~2 and~3 until quenched by dipole--dipole (rate $k_\mathrm{DD}$) and molecule--wall collisions (rate $k_\mathrm{w}$). }
\label{figS_3level}
\end{figure}

Figure~\ref{figS_3level}(a) schematically illustrates the QPML system. After passing through the front mirror, the QCL pump beam excites the molecular gain medium within the copper-tube laser cavity of radius $R_\mathrm{cell}$ and length $L$. The rotational lasing transition of the gain molecule is brought into resonance with the cavity by a back mirror that adjusts the cavity length. The terahertz laser power emerges from the same front mirror with transmission coefficient $T_\mathrm{THz}$ (hereafter referred to as $T$ unless specified otherwise), after which it is detected by a Schottky diode detector or heterodyne receiver.  We envision the output coupler as a copper reflector with a centered pinhole through which both beams pass; however, the analysis presented here does not require this geometry, only that the cavity's infrared and terahertz input/output transmissivities are known.

Throughout the paper, N$_2$O is adopted as the gain molecule because of the simplicity of its rotational energy-level structure, large infrared absorption coefficients, and experimentally demonstrated performance as a broadly tunable QPML~\cite{Science2019}. However, the revised three-level model presented below, and the design principles derived from it, are universal and may be easily applied to other candidate linear molecules such as HCN, OCS, and CO.  Extension of this model to more complex QPML molecules such as NH$_3$ and CH$_3$F may also be made, requiring the inclusion of additional collisional relaxation processes and rates, but the results derived herein may be applied to those molecules without loss of generality.  As has been discussed previously~\cite{Science2019}, the tuning range, pump threshold power, and output power depend on the rotational energy-level spacings of the molecule, its infrared absorption coefficients, its dipole moment, and the gas pressure $p$.  What has not been explored, and will be discussed extensively below, is how this also depends on pump power, pump saturation, cavity geometry $R_\mathrm{cell}$ and $L$, and output-coupler transmissivity $T$. 

In the following sections, we will first provide an overview of the key physical processes that must be captured in the model. Then we
describe in detail the rate equation for obtaining the population inversion, pump saturation, gain coefficient, and the approach for computing the pump threshold power, the output power, and the laser efficiency. We show that results obtained from the three-level model agree well with the comprehensive model, with peak power difference less than 15\%. Then a full discussion on the cavity design will be presented focusing on the cavity loss, cavity dimensions, and output coupler transmission over different laser transitions. 

\section{Laser-Model Overview}

\rev{A theoretical model describing the behavior of a QPML must account for the molecular collisional relaxation mechanisms that operate among the many thermally populated rotational states and act to quench the QCL-pumped inversion.  There are dozens of such thermally populated rotational states in the ground vibrational manifold of N$_2$O, and dozens more in the excited vibrational level when pumped by the QCL. This suggests that hundreds of collisional thermalization channels exist among those levels and are faithfully represented in the previously developed comprehensive model that describes how QPML power and threshold depend on pump power, pressure, and cavity parameters~\cite{Wang2018}.  However, in the very low pressure regime where molecular collisions with the wall of the laser cavity are more frequent than inter-molecular collisions, only the rotational states directly connected by the QCL pump or QPML transition experience significant pump-induced population changes. Consequently, it has been shown that a simple three-level model captures the salient QPML dependence on molecular and cavity parameters in this very low pressure regime~\cite{Science2019}. The improved three-level model introduced here extends its applicability to higher pressures by explicitly including the effects of inter-molecular collisions and pump saturation. It's viability as a descriptive model will be ascertained by comparing its predictions to those of the comprehensive model.}

Figure~\ref{figS_3level}(b) depicts the three-level model proposed in this work.  The population of these levels is partitioned into thermal and non-thermal contributions, respectively representing the unpumped and pumped molecules whose rotational populations can or cannot be represented by a Boltzmann distribution.  $N_1$, $N_2$, and $N_3$ are the non-thermal populations induced by the pump for the three rotational levels. The total population for level $i$ is the summation over thermal and non-thermal populations $N^\mathrm{tot}_i = N_i + N_i^{(0)}$, where $N_i^{(0)}$ is the thermal population without the pump. The QCL excites a ro-vibrational transition from rotational level~1 with quantum number~$J_L$ in the ground vibrational state to rotational level~2 in the excited vibrational state~v$_3$ in N$_2$O. The rotational quantum number of level~2 is determined by the type of ro-vibrational transition excited by the QCL: for $P$-, $Q$-, and $R$-branch transitions, $J_2$ = $J_L-1$, $J_L$, and $J_L+1$, respectively. Without loss of generality, we only consider the $R$-branch excitation here, so the rotational quantum numbers of levels~2 and~3 are~$J_L+1$ and~$J_L$, respectively.  
By using a continuously tunable mid-infrared QCL, every thermally populated ground state rotational level~$J_L$ may be pumped, inducing the corresponding v$_3$ rotational transitions $J_L+1\to J_L$ to lase, making broad terahertz tunability a reality~\cite{Science2019}. (This paper does not address lasing arising from the ground-state refilling transitions $J_L+1\to J_L$.) The laser frequency is $\nu_\mathrm{THz} \approx 2B(J_L+1)$, where $B$ is the rotational constant. For N$_2$O, $2B\approx 24.9$~GHz.

The dominant collisional-relaxation processes for the pressure range considered here ($< 1$~Torr) are the dipole--dipole and molecule--wall collisions. For linear molecules, the branching ratio for rotational transitions following dipole--dipole collisions is $\rho_{J\to J+1} = (J+1)/(2J+1)$, and $\rho_{J\to J-1} = J/(2J+1)$. For $J>10$, we can approximate $\rho_{J\to J+1} \approx \rho_{J\to J-1} \approx 1/2$, and hence the dipole--dipole collision rate is $k_\mathrm{DD}/2$ as shown in Fig.~\ref{figS_3level}. Here, $k_\mathrm{DD} = \sigma_\mathrm{DD} n_\mathrm{tot}\langle v\rangle$, where $\sigma_\mathrm{DD}=35\mathrm{\AA}^2$ is the dipole--dipole collision cross section for N$_2$O, $n_\mathrm{tot}$ is the total population density (proportional to the molecular pressure), and $\langle v \rangle$ is the thermally averaged relative velocity between molecules. The other half of the dipole--dipole collisions (up from level~2 and down from level 3) transfer population into the v$_3$ vibrational ``thermal pool''~\cite{McCormick1987,Everitt1986,Wang2018}, a collection of rotational levels whose total population may rise and fall in response to the action of the pump but whose relative rotational population distribution remains in thermal equilibrium with a Boltzmann distribution. As noted above, the thermal pool adds thermally distributed population into levels~2 and~3 that tends to quench the QCL-induced non-thermal population inversion. However, as will be discussed below, the non-thermal population in N$_2$O dominates and the thermal population is negligible. 

Molecule--wall collisions are also an important pathway for relaxation, especially in a compact laser cavity. In the low-pressure limit, the molecular mean free path $\lambda_M$ is larger than the \rev{cylindrical-cavity radius $R_\mathrm{cell}$}, so pressure-independent ballistic collisions with the wall are dominant. At higher pressures where $\lambda_M  < R_\mathrm{cell}$, molecules diffuse to the walls at a rate that decreases inversely with pressure. \rev{The wall-collision rate used here is the minimum of the ballistic ($k_w = 2\langle v\rangle_\mathrm{abs} /3R_\mathrm{cell}$) and diffusive ($k_w = 2\langle v\rangle_\mathrm{abs} \lambda_M/3R_\mathrm{cell}^2$) wall rates,} where $\langle v\rangle_\mathrm{abs}$ is the average absolute velocity of the molecule, and the factor of $2/3$ is a geometrical factor accounting for the radial velocities in the cylindrical cavity~\cite{EverittThesis}. 

For N$_2$O, numerous vibrational levels lie below and above the excited v$_3$ level pumped by the QCL. Vibrational-state changing (or ``v--v") collisions among the thermal pools of these levels are another significant relaxation process, especially at higher pressures when their rate, which grows linearly with pressure, exceeds the rate of wall collisions. These v--v collisions represent a potential advantage of N$_2$O over molecules like CH$_3$F and NH$_3$ which have no intermediate vibrational levels between the ground and pumped vibrational levels. Consequently, as pressure increases and the diffusive wall collision rate slows, pumped population accumulates in the thermal pool and quenches the inversion~\cite{Everitt1986,Wang2018}. 
Although this ``vibrational bottleneck" is negligible over the range of low pressures in which the N$_2$O laser operates, these additional v--v relaxation pathways need to be included in the model at very high pressures or for other molecules such as CH$_3$F and NH$_3$.

Note that spontaneous emission from rotational levels is neglected when computing the population inversion in our three-level model. In these long-wavelength, collision-dominated molecular gas lasers, the spontaneous-emission rate ($1/t_\mathrm{sp}\sim10^{-5}s^{-1}$) is negligible compared with the dipole--dipole collision rate ($k_\mathrm{DD}\sim 10^{5}s^{-1}$). Therefore, the spontaneous-emission rate between levels~2 and~3 is used only for the gain calculation.

Of paramount importance for understanding lasing in compact cavities is the effect of pump saturation on the QCL-pumped infrared ro-vibrational transition.  As the pump moves significant population from level~1 to level~2, the infrared-absorption coefficient $\alpha_\mathrm{IR}$ decreases as $N_1 - N_2$ decreases.  Since the QCL pump decays along the cavity axis exponentially as $\exp{(-\alpha_\mathrm{IR} L)}$, a decreasing $\alpha_\mathrm{IR}$ means an increasingly transparent gas and a greater penetration depth for the QCL pump. This pump-saturation effect, which is greatest for low-pressure operation in short, small-diameter cavities, depends on the pump rate, which also depends on $\alpha_\mathrm{IR}$. Consequently, the nonlinear effects of pump saturation must be ascertained self-consistently using an iterative nonlinear-equation algorithm.

\section{Three-level Model Equations}

The rate equations for $N_i$ as shown in Fig.~\ref{figS_3level}(b) are
\begin{equation}
    \begin{aligned}
        \frac{dN_1}{dt} &= -R_\mathrm{pump} - N_1(k_\mathrm{DD}+k_w),\\
        \frac{dN_2}{dt} &= R_\mathrm{pump} - N_2(k_\mathrm{DD}+k_w) + N_3 k_\mathrm{DD}/2,\\
        \frac{dN_3}{dt} &= N_2 k_\mathrm{DD}/2 - N_3(k_\mathrm{DD}+k_w).
    \end{aligned}
    \label{eq:rateeq}
\end{equation}
At steady state, $dN_1/dt = dN_2/dt = dN_3/dt = 0$ so that
\begin{equation}
    \begin{aligned}
    N_1 &= -\frac{R_\mathrm{pump}}{k_\mathrm{DD} + k_w},\\
    N_2 &= R_\mathrm{pump}\frac{k_\mathrm{DD}+k_w}{(3k_\mathrm{DD}/2 + k_w)(k_\mathrm{DD}/2 + k_w)},\\
    N_3 &= R_\mathrm{pump}\frac{k_\mathrm{DD}/2}{(3k_\mathrm{DD}/2 + k_w)(k_\mathrm{DD}/2 + k_w)},
    \end{aligned}
    \label{eq:steadystate}
\end{equation}
and the population inversion between levels~2 and~3 is
\begin{equation}
    \Delta N = N_2 - N_3 = \frac{R_\mathrm{pump}}{3k_\mathrm{DD}/2 + k_w}.
    \label{eq:populationinversion}
\end{equation}
Here, the pump rate \rev{for QCL power $P_\mathrm{QCL}$} is
\begin{equation}
    R_\mathrm{pump} = \frac{\beta P_\mathrm{QCL}}{h \nu_\mathrm{IR}} \frac{1}{\pi R_\mathrm{cell}^2 L} 
    \label{eq:r_pump}
\end{equation}
in which $\beta$ is the fraction of pump power absorbed in the cavity. $\beta$ depends on $P_\mathrm{QCL}$, pressure, $\alpha_\mathrm{IR}$ and $L$ in a complex manner. In the low-pressure limit with consideration of only a single pass in the cavity as in Ref.~\cite{Science2019}, $\beta \approx \alpha_\mathrm{IR}L$ $\ll$ 1. More generally, the QCL beam reflects from the back and front mirrors and, over multiple bounces, can travel much farther than $L$. As pressure increases, so does $\alpha_\mathrm{IR}$, effectively reducing the number of bounces by the QCL beam. For increasing pressure, $\alpha_\mathrm{IR}$, or $L$, $\beta$ approaches 1, so the entire pump beam is absorbed. To capture these effects, including pump saturation and round-trip bounces and losses, a more accurate description of the pump absorption is needed.

With a given $\alpha_\mathrm{IR}$, $\beta$ can be obtained by summing contributions over all round-trips as
\begin{equation}
    \beta = \left(1-e^{-\alpha_\mathrm{IR} L}\right)\frac{ \left(1+R_1 e^{-\alpha_\mathrm{IR} L}\right)}{1-R_1R_2 e^{-2\alpha_\mathrm{IR} L}}
    \label{eq:beta}
\end{equation}
where $R_1$ and $R_2$ are the IR reflection coefficients from the back and front mirrors, respectively. Notice that $\beta$, which is simply the Beer's law absorption multiplied by a term that accounts for multiple round trips, includes both molecular ($\alpha_\mathrm{IR}$, pressure) and cavity ($R_1$, $R_2$, $L$) contributions. Thus, the fractional pump power absorbed depends on and may be adjusted by any of these parameters. For a copper mirror, we approximate $R_1= 0.95$ and $R_2 = T_\mathrm{IR} R_1$, with $T_\mathrm{IR}$ = 0.96 (4\% leaking out the front mirror).
Since $\beta$ in \eqref{eq:beta} determines the pump rate~\eqref{eq:r_pump} and the populations $N_1$ and $N_2$~\eqref{eq:steadystate}, it affects the IR pump absorption coefficient $\alpha_\mathrm{IR}$ nonlinearly.
$\alpha_\mathrm{IR}$ at the pump frequency $\nu_\mathrm{IR}$ is evaluated by
\begin{equation}
    \alpha_\mathrm{IR} = \frac{8\pi^3 \nu_\mathrm{IR}}{3hc} |\langle 1 |\mu|2\rangle|^2 (N_1^\mathrm{tot}-N_2^\mathrm{tot})S(\nu_\mathrm{IR},\nu_0),
    \label{eq:alphair}
\end{equation}
in which $N_1^\mathrm{tot} = N_1^{(0)} + N_1$ and $N_2^\mathrm{tot} = N_2^{(0)} + N_2$ are the total (thermal + non-thermal) populations of levels 1 and 2, $|\langle 1 |\mu|2\rangle|^2$ is the dipole matrix element for the pump transition obtained from the HITRAN database~\cite{Gordon2017}, and $S(\nu_\mathrm{IR},\nu_0)$ is the absorption lineshape function, obtained by convolving the spectral hole burning (SHB) and the Gaussian Doppler lineshape functions. For IR absorption, the SHB lineshape is a Lorentzian obtained by convolution of two other Lorentzians: the pressure broadening lineshape with half width $\Delta \nu_p$ and the QCL linewidth with half width $\Delta\nu_\mathrm{QCL}$, summed together as $\Delta\nu_\mathrm{SHB} = \Delta\nu_p + \Delta\nu_\mathrm{QCL}$. $S(\nu_\mathrm{IR},\nu_0)$, which is usually described by a Voigt profile, is empirically approximated here by a Lorentzian with width $\Delta\nu_S = \sqrt{\Delta\nu_D^2 + \Delta \nu_\mathrm{SHB}^2}$. This approximation underestimates the effect of pump saturation at very low pressures, but as $\beta \to 1$ all pump photons are absorbed irrespective of the lineshape approximation used. For N$_2$O, $\Delta \nu_p =4$~MHz/Torr, and $\Delta\nu_D = 3.58\times10^{-7}\nu_\mathrm{0}\sqrt{T_e/M}$ in which $\nu_0$ is the IR transition frequency (both MHz), $T_e=300K$ is the temperature, and $M=44$ amu is the molecular mass. The linewidth of the pump QCL is assumed as $\Delta\nu_\mathrm{QCL}=1$~MHz~\cite{Science2019}.


\begin{figure}[htp]
\centering
\includegraphics[width=0.45\textwidth]{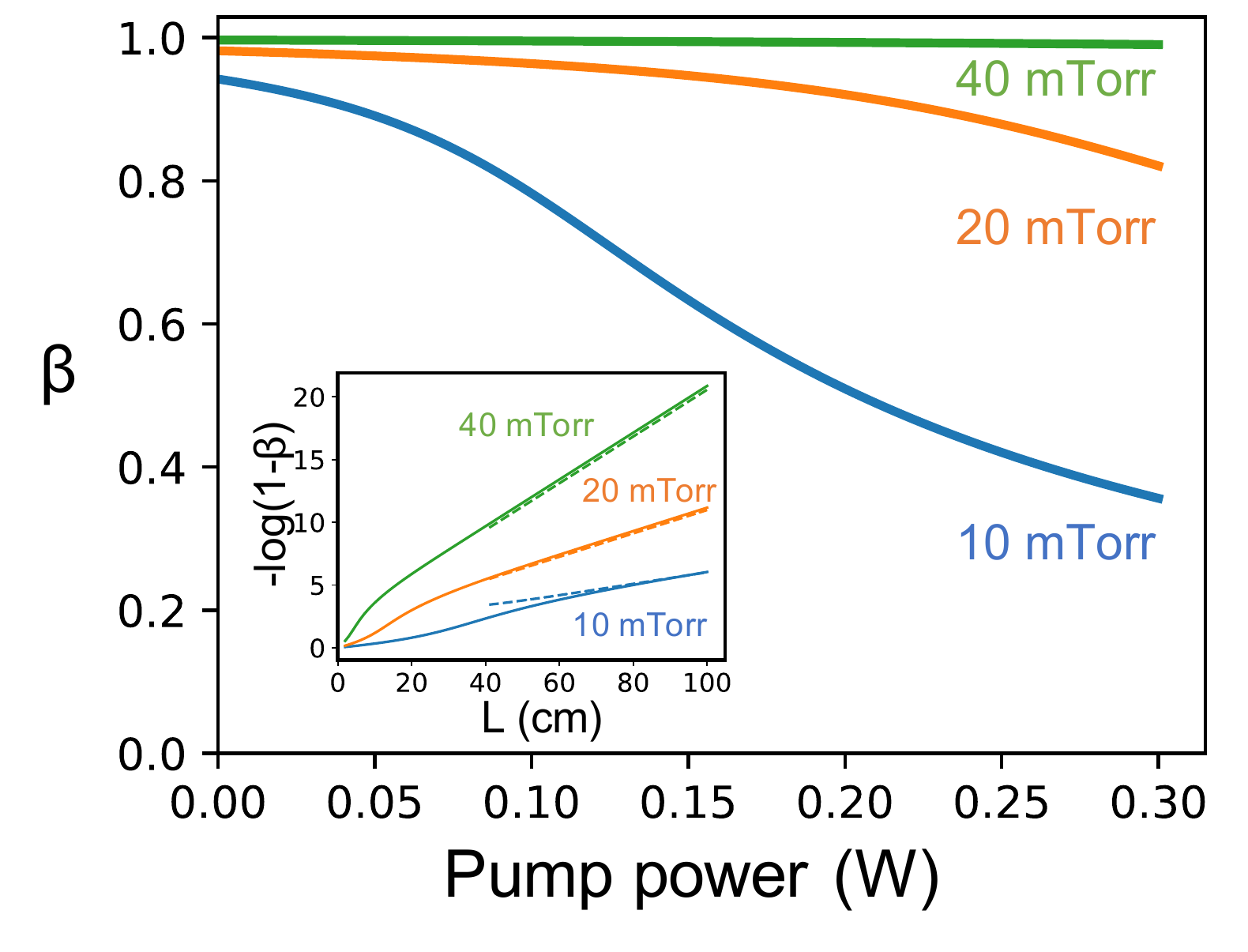}
\caption{The fraction of absorbed IR pump $\beta$ varies with pump power, for three different molecular pressures. In the calculation,  rotational state $=14$ is pumped with power 250~mW, and the cavity geometry is $R_\mathrm{cell} = $0.25~cm, and $L = 15$~cm. The inset shows how IR absorption changes with cavity length $L$ for different molecular pressures. Dashed lines plot $(\alpha_\mathrm{IR} L - \log(1-R_1))$ for the three pressures, showing an asymptotic linear relation between $(-\log(1-\beta))$ and $L$ for longer cavities.}
\label{fig:beta}
\end{figure}


The saturation-dependent infrared absorption coefficient for a given transition is obtained by nonlinearly solving equations \eqrefs{eq:populationinversion}{eq:alphair}. Pump saturation is described in \eqref{eq:alphair} by the population difference $(N_1^\mathrm{tot} - N_2^\mathrm{tot})$ between levels~1 and~2. As mentioned above, level~1 depletes and level~2 fills as pump power increases so that $(N_1^\mathrm{tot} - N_2^\mathrm{tot})$ and $\alpha_\mathrm{IR}$ decrease. 
Figure~\ref{fig:beta} plots typical fractional IR absorption curves as a function of pump power for three N$_2$O pressures. As expected, $\alpha_\mathrm{IR}$ decreases as the pump power increases, allowing more round trips of the IR beam inside the cavity. This pump saturation is more significant at lower pressures because there is less population in level~1 and the SHB is narrower.  It can be seen that saturation increases almost quadratically with decreasing pressure. Note that some infrared power is lost on each round trip through absorption by the cavity walls and through the output coupler, so less of the infrared beam is absorbed by the gas in multiple round trips than in a cavity long enough for single-pass absorption. Indeed, in the limit of an empty cavity, the IR beam may only execute $(1+R_1)/(1-R_1 R_2)$ round trips (14.6 for the values of $R_1$ and $R_2$ presented above) before being fully lost by the cell itself. The inset illustrates these behaviors by plotting the relation between fractional loss ($-\log(1-\beta)$) and the cavity length $L$ for a typical pumped transition. Without pump saturation and round trips in the cavity, one expects from \eqref{eq:beta} a linear curve $-\log(1-\beta)\approx \alpha_\mathrm{IR} L + \log(1-R_1)$ whose slope is the IR absorption coefficient that increases linearly with pressure and $L$ for $L >$ 30~cm. The dashed lines in the inset of \figref{fig:beta}  plot $(\alpha_\mathrm{IR} L - \log(1-R_1))$ for the three pressures, confirming this asymptotic linear relation. For shorter cavities, pump saturation and multiple round trips produce larger cavity losses and reduce the fraction of pump power $\beta$ absorbed by the molecular gain medium. 

Note that in the three-level model, molecules transfer from levels~2 and~3 into the thermal pool with rate $k_\mathrm{DD}/2$, as shown in \figref{figS_3level}. For N$_2$O, the thermal pool relaxes to the ground state with rate $k_\mathrm{pool} = k_w + k_\mathrm{vv}$ through wall and v--v collisions. To obtain the v--v transition rate $k_\mathrm{vv}$, we need to estimate the probability of a collision hard enough to supply the v--v energy difference $\Delta E$ assuming a Boltzmann velocity distribution $\sigma_\mathrm{vv} = \sigma_\mathrm{GKC} \exp(-\Delta E/k_BT_e)$, where $\sigma_\mathrm{GKC}=15\AA^2$ is the gas kinetic cross section. For N$_2$O, the zero-point energy of the vibrational level just below v$_3$ is approximately 350~cm$^{-1}$ lower, giving $\sigma_\mathrm{vv}\approx 0.2\sigma_\mathrm{GKC}$. But for molecules like CH$_3$F, v$_0$ is the only vibrational level below v$_3$, so $\sigma_\mathrm{vv}$ is negligibly small ($\approx 0.005\sigma_\mathrm{GKC}$ for that 1050~cm$^{-1}$ energy separation).

Using \eqref{eq:steadystate} and \eqref{eq:populationinversion}, the thermal-pool population may be estimated as
\begin{equation}
    \begin{split}
    N_p & = \frac{k_\mathrm{DD}}{2k_\mathrm{pool}} (N_2+N_3) \\
    & = \frac{(3k_\mathrm{DD}/2+k_w)}{(k_\mathrm{DD}/2+k_w)} \frac{k_\mathrm{DD}}{2k_\mathrm{pool}} \Delta N\\
    & =\tilde{k} \Delta N.
    \end{split}
\end{equation}
 The thermal population added to levels~2 and~3 acts to quench the population inversion by $(f_3-f_2)N_p=(f_3-f_2)\tilde{k}\Delta N$, where $f_2$ and $f_3$ are the respective thermal population fractions. Because $k_\mathrm{vv} \gg k_w$ for all but the smallest diameter cavities and lowest pressures in N$_2$O, $\tilde{k}\lesssim 20$, and since $f_3~-~f_2~<~10^{-3}$ for $J_L>10$, the thermal quenching effect on $\Delta N$ is negligible. Therefore, $N_p$ may be excluded from the three-level model for N$_2$O.

But the thermal pool population becomes important in other molecules when vibrational relaxation mechanisms are slow and population accumulates in the pumped vibrational level, producing vibrational bottlenecking. In such cases, $k_\mathrm{vv}\to0$, and $\tilde{k} \to 3k_\mathrm{DD}/2k_\mathrm{pool}=3k_\mathrm{DD}/2k_w$ which increases quadratically with pressure considering $k_\mathrm{DD}\propto p$ and $k_w\propto 1/p$. Fortuitously, as thermal population accumulates in the excited thermal pool, the vibrational bottleneck effect is somewhat circumvented at high pressures through additional v--v collisions involving higher lying vibrational levels, an effect which can be simulated by introducing an effective vibrational temperature representing the population of an ensemble of many high-lying vibrational levels~\cite{Everitt1986,Wang2018}.

Using the calculation of the pump absorption, pump saturation, and population inversion, one may obtain the gain coefficient $\gamma$, pump threshold $P_\mathrm{th}$, conversion efficiency $\eta$, and output power $P_\mathrm{THz}$ of the QPML. The unsaturated gain coefficient~\cite{STbook} is 
\begin{equation*}
    \gamma_0 = \sigma \Delta N 
\end{equation*}
where $\sigma = {\lambda_\mathrm{THz}^2}/{8\pi^2 \Delta \nu t_\mathrm{sp}}$ is the cross section for the lasing transition, the spontaneous emission lifetime is $t_\mathrm{sp} = 3h\epsilon_0\lambda_\mathrm{THz}^3/16 \pi^3 \mu_{23}^2$, where $\mu_{23}^2 = | \langle 2 | \mu | 3\rangle |^2$ is the dipole matrix element for the laser transition, and $\Delta \nu$ is the half width of the gain profile, approximated as $\Delta \nu \approx \sqrt{\Delta \nu_D^2+\Delta\nu_p^2}$, where $\Delta\nu_D$ and $\Delta\nu_p$ are the Doppler broadening (for the terahertz transition) and pressure broadening linewidths, respectively. 

The laser threshold occurs when the gain coefficient balances the total cavity loss $\alpha_\mathrm{cell}$ at the terahertz laser frequency,
\begin{equation}
    \gamma_0 = \alpha_\mathrm{cell}.
\end{equation} 
The total cavity loss consists of the terahertz transmission loss through the front mirror and internal Ohmic losses. The latter can be computed analytically for the modes of a hollow metal waveguide~\cite{ohmicloss}, and we assume throughout this paper that the cylindrical QPML lases in the lowest-loss TE$_\mathrm{01}$ mode.  From this, the threshold power can be obtained as
\begin{equation}
    P_\mathrm{th} = \frac{h \nu_\mathrm{IR}}{\sigma}\frac{\alpha_\mathrm{cell}}{\beta}(\pi R_\mathrm{cell}^2 L)(1.5k_\mathrm{DD} + k_w) .
    \label{eq:pth}
\end{equation}

\begin{figure}[h]
\centering
\includegraphics[width=0.5\textwidth]{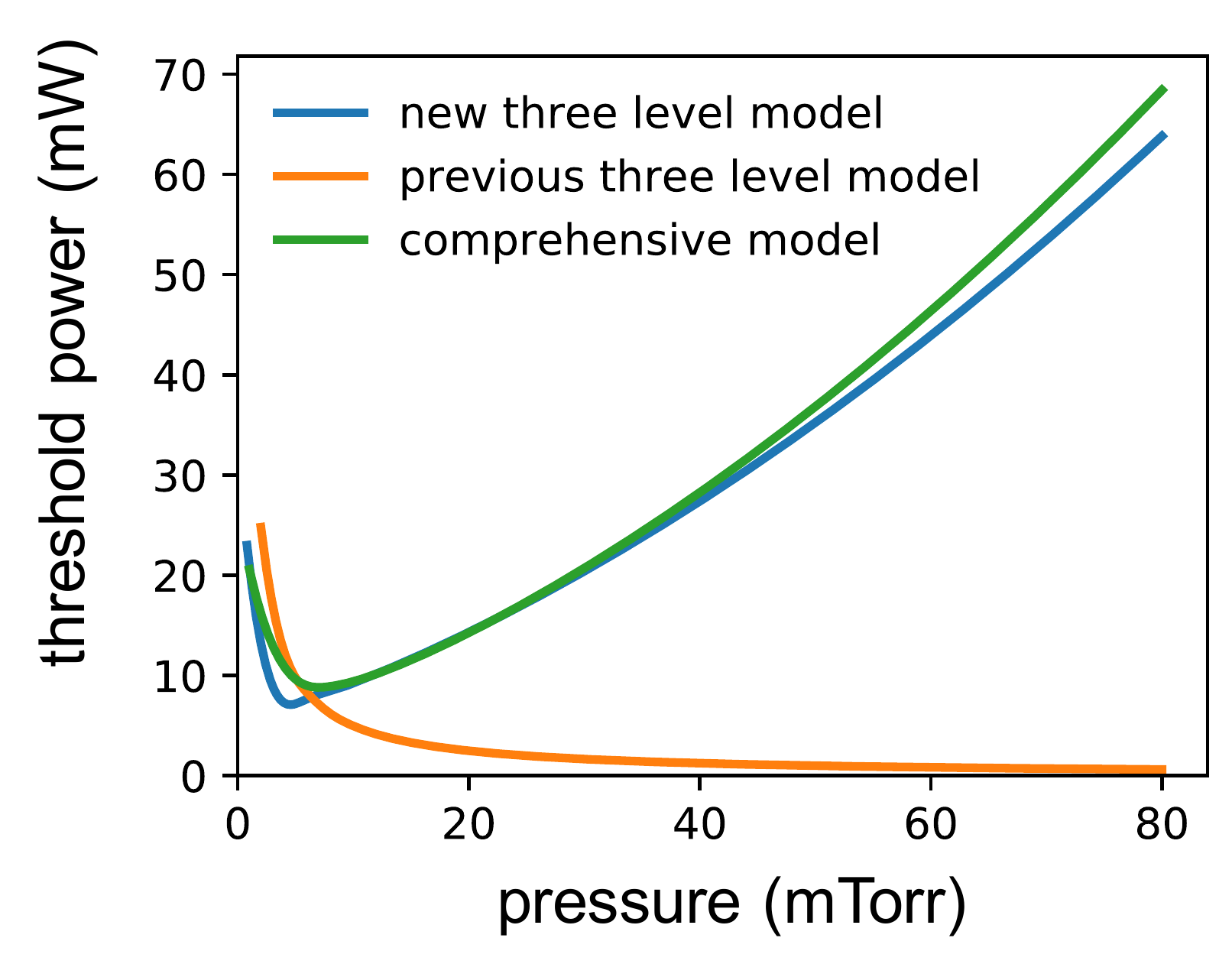}
\caption{Pump threshold $P_\mathrm{th}$ as a function of pressure, compared with the prior three-level model~\cite{Science2019,chevalier2020response} and the comprehensive model~\cite{Wang2018}, for a cylindrical copper cavity with $R_\mathrm{cell}=0.25$~cm, $L=15$~cm, and $T=0.016$ filled with N$_2$O is pumped by a QCL with $\Delta\nu_\mathrm{QCL}=1$~MHz coincident with the $R$-branch IR transition $J=14,\mathrm{v}_0 \to J=15,\mathrm{v}_3$.}
\label{fig:Pth}
\end{figure}

Figure~\ref{fig:Pth} compares the pump threshold as a function of pressure with that calculated using the comprehensive numerical simulation~\cite{Wang2018} and the prior three-level model~\cite{Science2019} for a cylindrical copper cavity with $R_\mathrm{cell}=0.25$~cm, $L=15$~cm, and $T=0.016$. In the calculations, the QCL with full linewidth 2~MHz excites the $R$-branch IR transition $(J=14,\mathrm{v}_0) \to (J=15,\mathrm{v}_3)$. At low pressures dominated by pressure-independent ballistic wall collisions, $\beta\propto \alpha_\mathrm{IR} L \propto p$ and $k_w \gg k_\mathrm{DD}$, so $P_\mathrm{th} \propto 1/p$. At higher pressures characterized by diffusive wall collisions (k$_w$ $\propto 1/p$) and $\beta \to$ 1, $P_\mathrm{th} \propto k_\mathrm{DD} \propto p$.  These behaviors are clearly observed in Figure~\ref{fig:Pth}, which shows how closely our new three-level model agrees with the comprehensive model across all pressures, while the simpler three-level model proposed in~\citeasnoun{Science2019,chevalier2020response} only works well in the low-pressure limit for which it was designed.  

The output power $P_\mathrm{THz}$ is computed with the condition that the saturated gain coefficient equals the cavity loss, i.e.,
\begin{equation}
    \gamma = \frac{\gamma_0}{1+\Phi/\Phi_s} = \alpha_\mathrm{cell}
\end{equation}
in which $\Phi$ is the photon flux density, $\Phi_s = k_s/\sigma$ is the saturated photon flux density, and $k_s$ is the saturated characteristic rate. In our three-level model, $k_s=(3k_\mathrm{DD}/2 + k_w)/2$ following the derivation in Ref.~\cite{STbook}. From this, the output power can be calculated as~\cite{STbook}
\begin{equation}
    \begin{aligned}
        P_\mathrm{THz} &=  \frac{T}{2} h\nu_\mathrm{THz} (\pi R_\mathrm{cell}^2) \Phi \\
        & = \frac{T}{2} h\nu_\mathrm{THz} (\pi R_\mathrm{cell}^2) \left( \frac{\gamma_0}{\alpha_\mathrm{cell}} - 1 \right)\frac{k_s}{\sigma}.
    \end{aligned}
    \label{eq:thzpower1}
\end{equation}

\begin{figure}[t]
\centering
\includegraphics[width=0.45\textwidth]{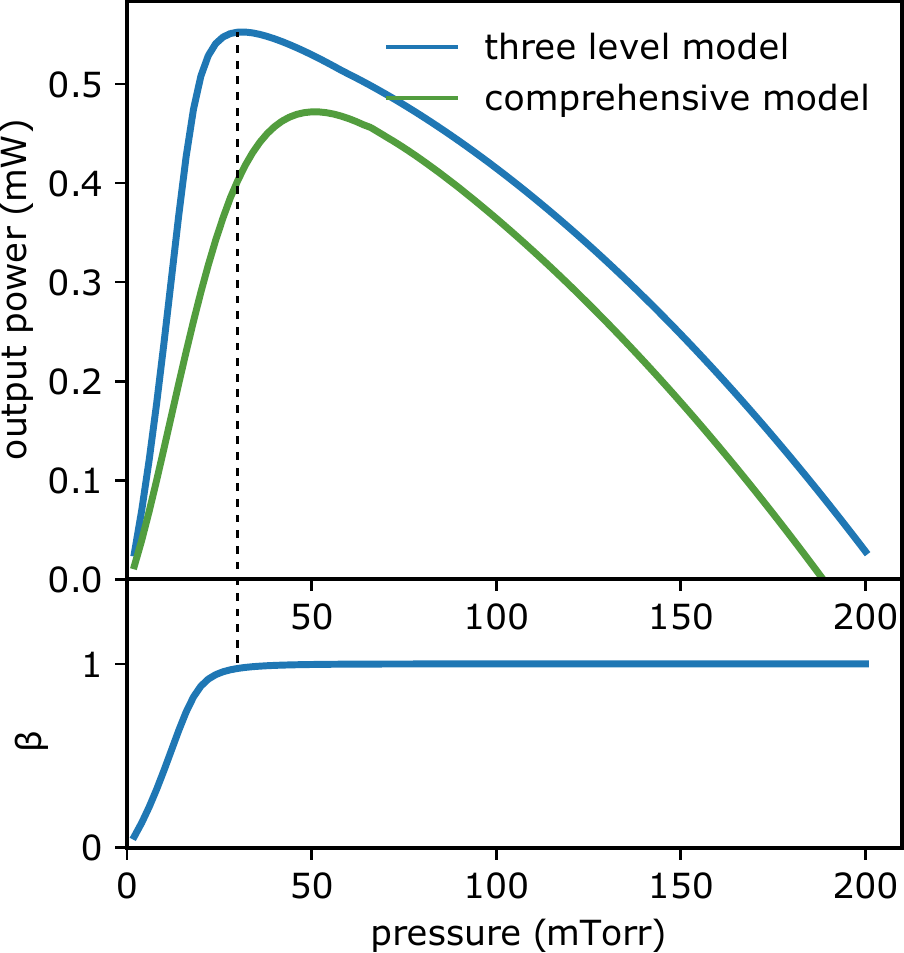}
\caption{Pressure-dependent QPML output power predicted by the three-level model and the comprehensive model. A 250mW QCL pumps the R-branch transition $J=14,\mathrm{v}_0 \to J=15,\mathrm{v}_3$ in a cavity of radius $R_\mathrm{cell}=0.25$~cm, length $L=15$~cm, and front mirror transmission $T=0.016$. The IR absorption $\beta$ from  the three-level model is plotted below to show the optimal pressure occurs as $\beta$ reaches 1.}
\label{fig:PTHz}
\end{figure}

Figure~\ref{fig:PTHz} plots $P_\mathrm{THz}$ as a function of pressure with 250mW of QCL power pumping the same transition considered in Figure~\ref{fig:Pth}. The three-level model's prediction quantitatively matches the comprehensive model to within 15\%, validating the methodology outlined above. (The three-level model slightly underestimated this optimal pressure because it slightly overestimated IR absorption at the low pressures and underestimated the threshold power (\figref{fig:Pth}) at higher pressures.) 

\section{Cavity design}

Thus, with the validated three-level model proposed in this paper, along with the comprehensive model~\cite{Wang2018}, let us now explore how $P_\mathrm{THz}$ depends on molecular and cavity parameters in order to explore optimal cavity designs and pumping strategies. Throughout, we will assume that one will use the most powerful QCL available and that the simplest way to increase QPML power is to increase QCL power. Consequently, we begin the discussion by assuming a fixed QCL power of 250~mW and explore how to optimize all other design and operational parameters.  \rev{Once the optimal cavity design is identified, we conclude by exploring how QPML power increases with increasing QCL pump power.}

\textbf{Pressure} is perhaps the easiest and most important operational parameter a user may adjust. The pressure dependence of $P_\mathrm{THz}$ is determined by the competing pressure dependencies of $\gamma_0$ and $k_s$. Both of these increase at low pressure, but at higher pressures $\Delta N$ and $\gamma_0$ shrink faster than $k_s$ grows, so an intermediate pressure produces maximal power. But underlying these behaviors is a helpful rule of thumb: the optimal pressure $p^\mathrm{opt}$ occurs as $\beta$ reaches 1 because at lower pressures the pump beam is not fully absorbed while at higher pressures the gas is not fully pumped.
This is illustrated in \figref{fig:PTHz} comparing the IR fractional absorption $\beta$ with the output power. 
Certainly these effects are exacerbated at low pressure by pump saturation and at high pressure by thermal quenching, but the primary factor determining maximum QPML power for a given transition and cavity geometry is the pressure at which absorption of the QCL pump is optimized. Consequently, calculating $\beta$ is critically important for estimating the performance of QPMLs. 

\begin{figure}[h]
    \centering
    \includegraphics[width=\columnwidth]{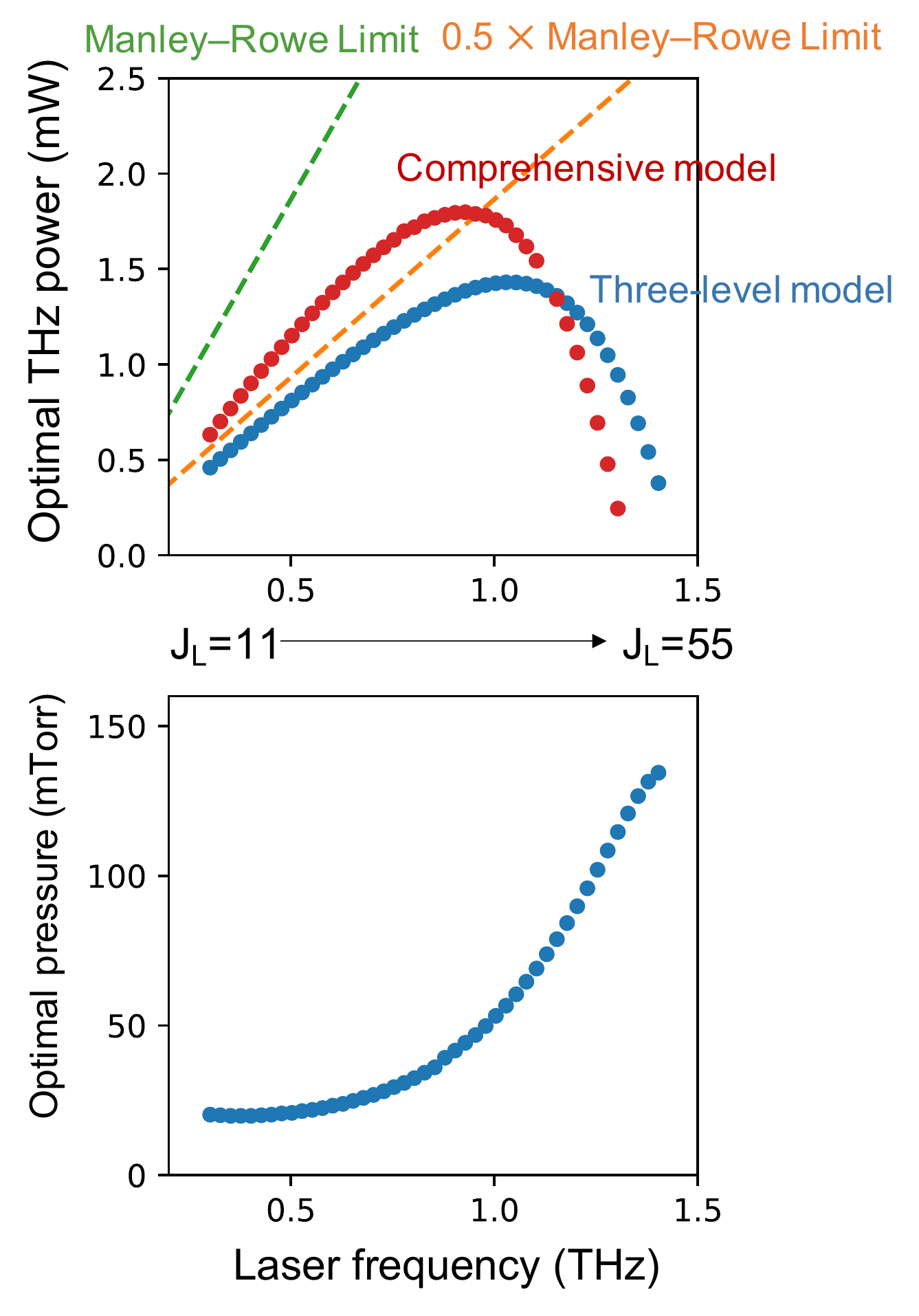}
    \caption{(a) Maximum output power vs laser frequency with $J_L=11$ to 51. (b) Optimal molecular pressure for (a) predicted by the three-level model for each laser frequency.}
    \label{fig:optmailPTHz_freq}
\end{figure}

\textbf{Frequency} of the QPML also affects the maximum power achievable. Of course, the QPML is a line-tunable laser, and the possible laser frequencies are the known rotational absorption transition frequencies of the molecular gain medium. The dependence of QPML power on this frequency may be seen if \eqref{eq:pth} is used to rewrite \eqref{eq:thzpower1} as
\begin{equation}
    \begin{aligned}
    P_\mathrm{THz} &= 
    \frac{\beta}{2}
    \left(\frac{h\nu_\mathrm{THz}}{h\nu_\mathrm{IR}}\right) 
     \left(\frac{T/2L}{\alpha_\mathrm{cell}}\right)
    (P_\mathrm{QCL} - P_\mathrm{th}),
    \end{aligned}
    \label{eq:thzpower2}
\end{equation}
with the power efficiency given by
\begin{equation}
    \eta = \frac{\beta}{2}
    \left(\frac{h\nu_\mathrm{THz}}{h\nu_\mathrm{IR}}\right) 
     \left(\frac{T/2L}{\alpha_\mathrm{cell}}\right).
     \label{eq:power_efficiency}
\end{equation}
The upper limit of the power efficiency $\eta$ is simply  
\begin{equation}
    \eta \le \frac{1}{2} \left(\frac{\nu_\mathrm{THz}}{\nu_\mathrm{IR}}\right),
\end{equation}
where the factor of $1/2$ (preceding the Manley--Rowe limit $\eta \leq \nu_\mathrm{THz} / \nu_\mathrm{IR}$) comes from the fact that $k_w$ is the same for levels~2 and~3. In the comprehensive model~\cite{Wang2018} with its larger number of energy levels and more comprehensive modeling of collisional transitions, this upper limit is somewhat relaxed, as shown in \figref{fig:optmailPTHz_freq}(a).

More importantly, since the pump frequency $\nu_\mathrm{IR}$ changes little over the tuning range of the molecular gain medium, the power efficiency and output power of a QPML increase with increasing $\nu_\mathrm{THz}$, as can be seen in \figref{fig:optmailPTHz_freq}(a). This  Manley--Rowe-related trend indicates that, for $P_\mathrm{QCL} \gg P_\mathrm{th}$ and all other factors being equal, higher power can be obtained at higher QPML frequencies as the energy efficiency of converting IR photons to terahertz photons increases. This is a tremendously important effect, because otherwise the terahertz frequency dependence of $P_\mathrm{THz}$ would mirror the IR frequency dependence of $\alpha_\mathrm{IR}$ for a fixed-length cavity.  Specifically, for N$_2$O, the $R$-branch transition with the largest $\alpha_\mathrm{IR}$ occurs at $J_L^\mathrm{max}$ = 15, corresponding to maximum $P_\mathrm{THz}$ occurring at only $\nu_\mathrm{THz}^\mathrm{max} \approx$ 0.399 THz. Instead, the Manley--Rowe effect pushes the transition that produces the most laser power much higher, toward $J_L \to 40$ and $\nu_\mathrm{THz}^\mathrm{max} \to$ 1 THz.

Since the population of $J_L$ decreases as $\nu_\mathrm{THz}$ increases above $J_L^\mathrm{max}$, the optimal pressure $p^\mathrm{opt}$ to achieve maximum laser power must increase with increasing $J_L$ (\figref{fig:optmailPTHz_freq}(b)). However, increasing pressure also increases $P_\mathrm{th}$ (\figref{fig:Pth}), so equation \eqref{eq:thzpower2} reveals how laser power drops precipitiously below the Manley--Rowe limit as $P_\mathrm{th}$ approaches  $P_\mathrm{QCL}$ with increasing $J_L$, thus establishing the high-frequency limit for laser operation.

\textbf{Cavity loss}, which deleteriously increases $P_\mathrm{th}$ \eqref{eq:pth} and decreases $P_\mathrm{THz}$ \eqref{eq:thzpower2}, is the next most important factor determining QPML power. Cavity loss is determined by two terms in $\alpha_\mathrm{cell} = \alpha_T + \alpha_o$, where $\alpha_o$ is the Ohmic loss of the metallic cavity and  
\eq{
\alpha_T = - \frac{\log(1-T)}{2L} \approx \frac{T}{2L}.
\label{eq:alpha_T}
}
is the transmission loss from the front mirror. For $T = 0.016$ and $L = 15$~cm, $\alpha_T \approx 0.053\,\mathrm{m}^{-1}$.

The Ohmic loss for the lowest-loss TE$_{01}$ mode in the cylindrical cavity is~\cite{ohmicloss}
\eq{
\alpha_o = \frac{R_s}{R_\mathrm{cell} \eta_0 \sqrt{1-(x_0/k_0R_\mathrm{cell})^2}} \left(\frac{x_0}{k_0 R_\mathrm{cell}}\right)^2,
\label{eq:alpha_o}
}
where $\eta_0=377~\Omega$ is the free-space impedance, $k_0=2\pi\nu_\mathrm{THz}/c$ is the wave number, $x_0=3.83$ is the Bessel-function zero for the TE$_{01}$ mode, and $R_s=\sqrt{\pi\nu_\mathrm{THz} \mu_0/\sigma_\mathrm{Cu}}$ is the RF sheet resistance (for which the conductivity of the copper cavity is $\sigma_\mathrm{Cu}$). For frequencies far above the waveguide cutoff, we can neglect the $\sqrt{1-(x_0/k_0R_\mathrm{cell})^2}$ term, and the Ohmic loss becomes
\eq{
\alpha_o = \left(\frac{x_0}{k_0}\right)^2 \frac{R_s}{\eta_0 R_\mathrm{cell}^3}.
}
Notice that Ohmic loss decreases with cell radius as $1/R_\mathrm{cell}^3$ and with laser frequency as $\nu_\mathrm{THz}^{-3/2}$, the latter contributing to the improved high-frequency performance of the laser. (However, cavity alignment becomes more challenging at high frequency, an effect not considered here as our calculation assumes a perfectly aligned cavity.) For the N$_2$O QPML, we have $\alpha_o\approx 0.00634/ (J_L+1)^{3/2}R_\mathrm{cell}^3 ~\mathrm{m}^{-1}$ with $R_\mathrm{cell}$ in cm.

\begin{figure}[h]
    \centering
    \includegraphics[width=\columnwidth]{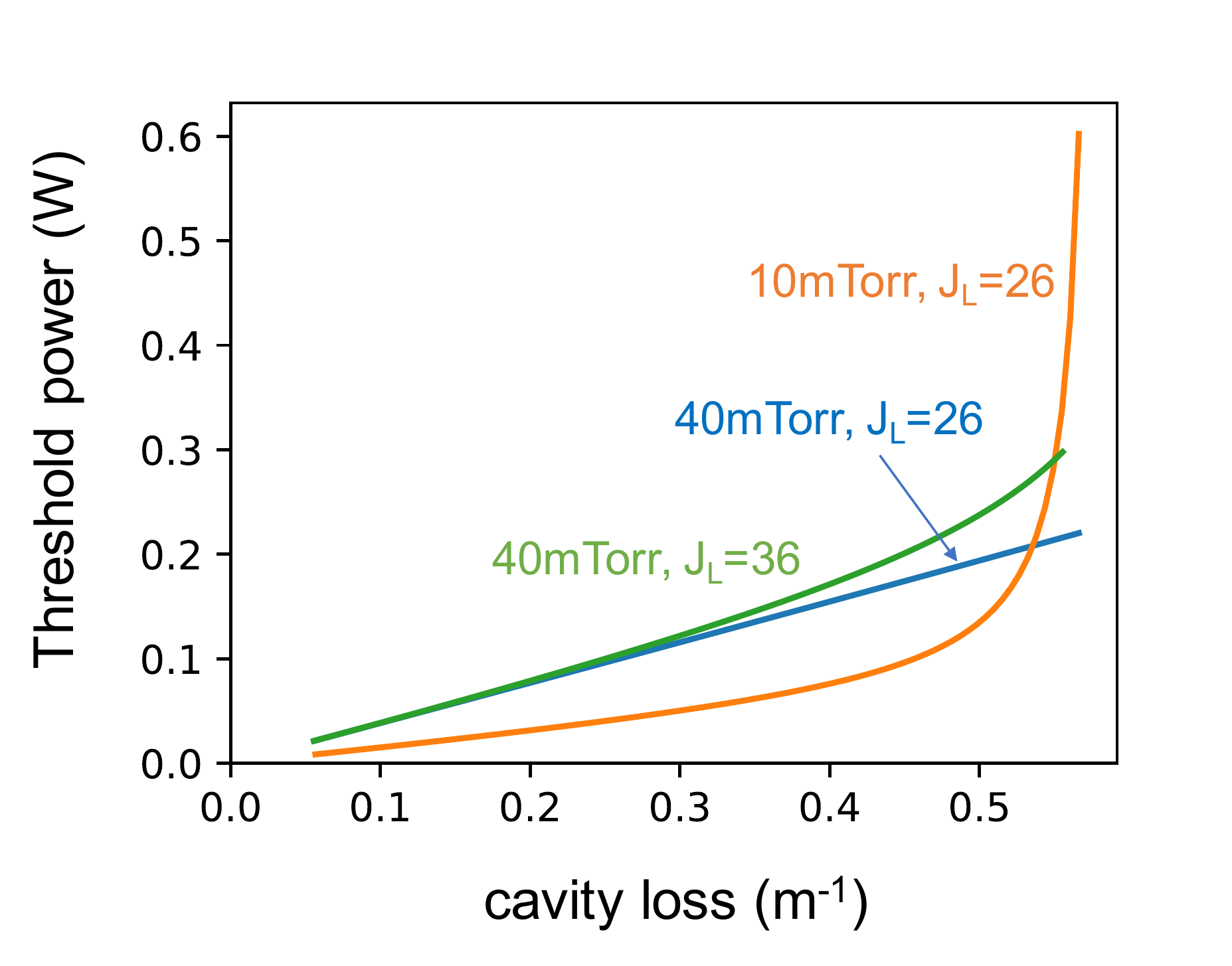}
    \caption{$P_\mathrm{th}$ as a function of total cavity loss for two transitions and two pressures.}
    \label{fig:Pth_cavityloss}
\end{figure}

Cavity loss $\alpha_\mathrm{cell}$ is usually dominated by transmission loss \eqref{eq:alpha_T}, which is much larger than Ohmic loss \eqref{eq:alpha_o} for typical compact cavities ($R_\mathrm{cell} > 0.1$~cm, $L < 1$~m, $T > 0.01$).  At first glance, Equation~\eqref{eq:pth} appears to imply a linear relation of threshold power $P_\mathrm{th}$ on $\alpha_\mathrm{cell}$. However, the relation is actually nonlinear because the cavity loss also affects $\beta$. \Figref{fig:Pth_cavityloss} shows how $P_\mathrm{th}$ increases with $\alpha_\mathrm{cell}$. The curves are linear at higher pressures as $\beta \to 1$ but become superlinear when pressure, $\alpha_\mathrm{IR}$, or $\beta$ are low. This again shows the importance $\beta$ and its accurate, self-consistent modeling for the cavity design. 

In the pressure regime for which $\beta \to 1$, which occurs at or above $p^\mathrm{opt}$, the output power \eqref{eq:thzpower2} from such a compact cavity may be approximated as 
\eq{
P_\mathrm{THz} \approx \frac{1}{2} \frac{\nu_\mathrm{THz}}{\nu_\mathrm{IR}} P_\mathrm{QCL}
\label{eq:PTHzShort}
}
when the pump power is much larger than the threshold power. The way $P_\mathrm{th}$ and $P_\mathrm{THz}$ depend on cavity parameters $R_\mathrm{cell}$, $L$, and $T$ will be addressed next.

\begin{figure}[h]
    \centering
    \includegraphics[width=\columnwidth]{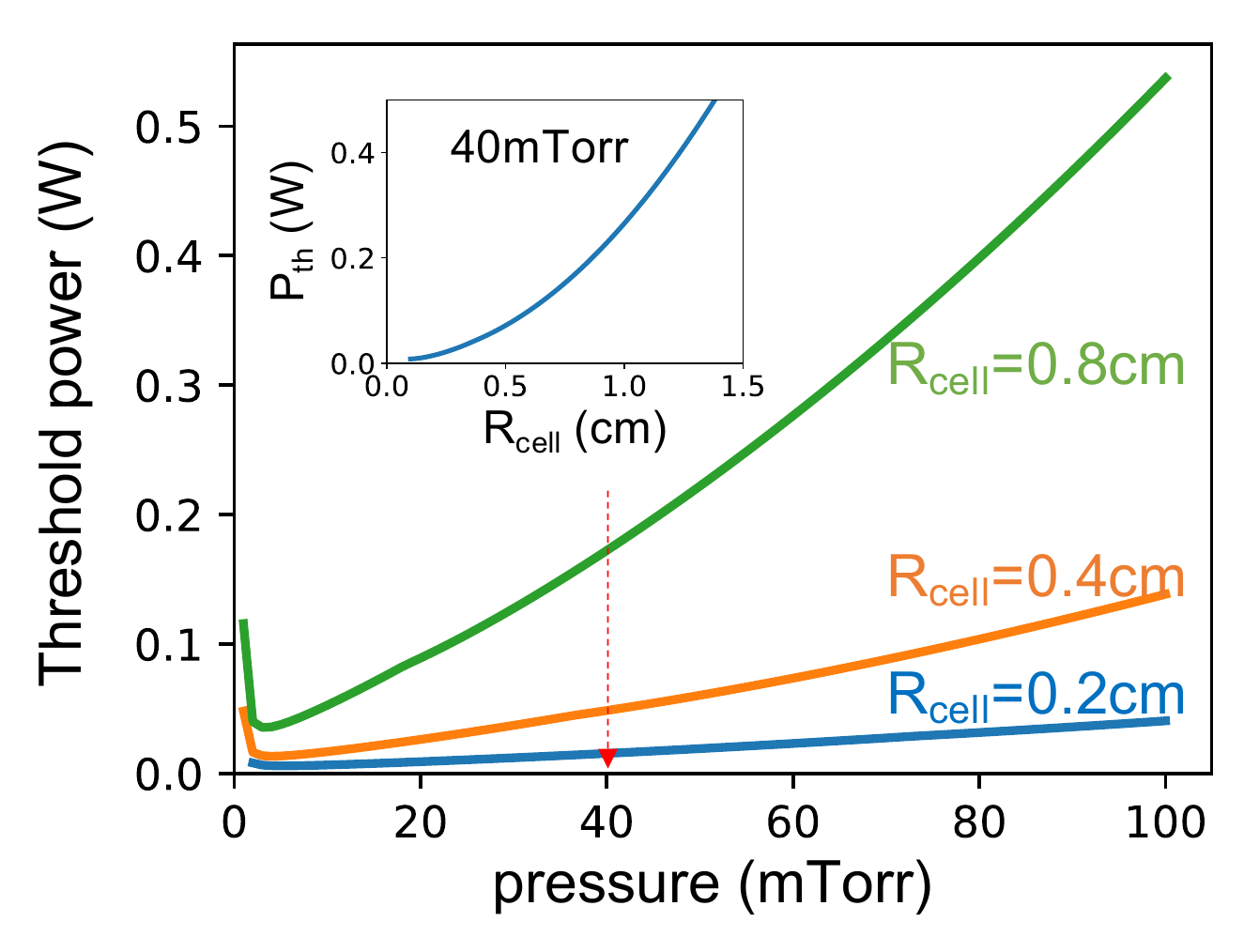}
    \caption{$P_\mathrm{th}$ plotted for QCL pump excitation of $J_L=26$ as a function of pressure for cavity radii $R_\mathrm{cell}=$0.2, 0.4, and 0.8~cm with fixed cavity length $L = 15$~cm and front mirror transmission $T=0.016$. The inset shows how $P_\mathrm{th}$ increases quadratically with $R_\mathrm{cell}$.}
    \label{fig:Pth_R}
\end{figure}

\textbf{Cavity Radius} primarily affects the threshold power. Equation \eqref{eq:pth} shows that for the compact cavities considered here with fixed $T$ and $\alpha_o \ll \alpha_T$, $P_\mathrm{th}$ increases quadratically with cavity radius. \Figref{fig:Pth_R} plots the threshold as a function of pressure for three different cavity radii. As described above, $P_\mathrm{th}$ increases quasi-linearly with pressure except at very low pressures, but notice how much $P_\mathrm{th}$ increases as cavity radius increases for a given pressure. This occurs because of the assumption that the QCL pump uniformly illuminates the cavity, so the pump rate must decrease with cavity cross section as $1/R_\mathrm{cell}^{2}$. The inset plots $P_\mathrm{th}$ as a function of $R_\mathrm{cell}$ with fixed pressure (40~mTorr) to confirm the quadratic relationship between lasing threshold and cavity radius. Of course, a larger radius will become detrimental in other ways, such as when the cavity becomes multi-moded or the pump does not spread enough to fill the whole diameter.

\begin{figure}[h]
    \centering
    \includegraphics[width=0.95\columnwidth]{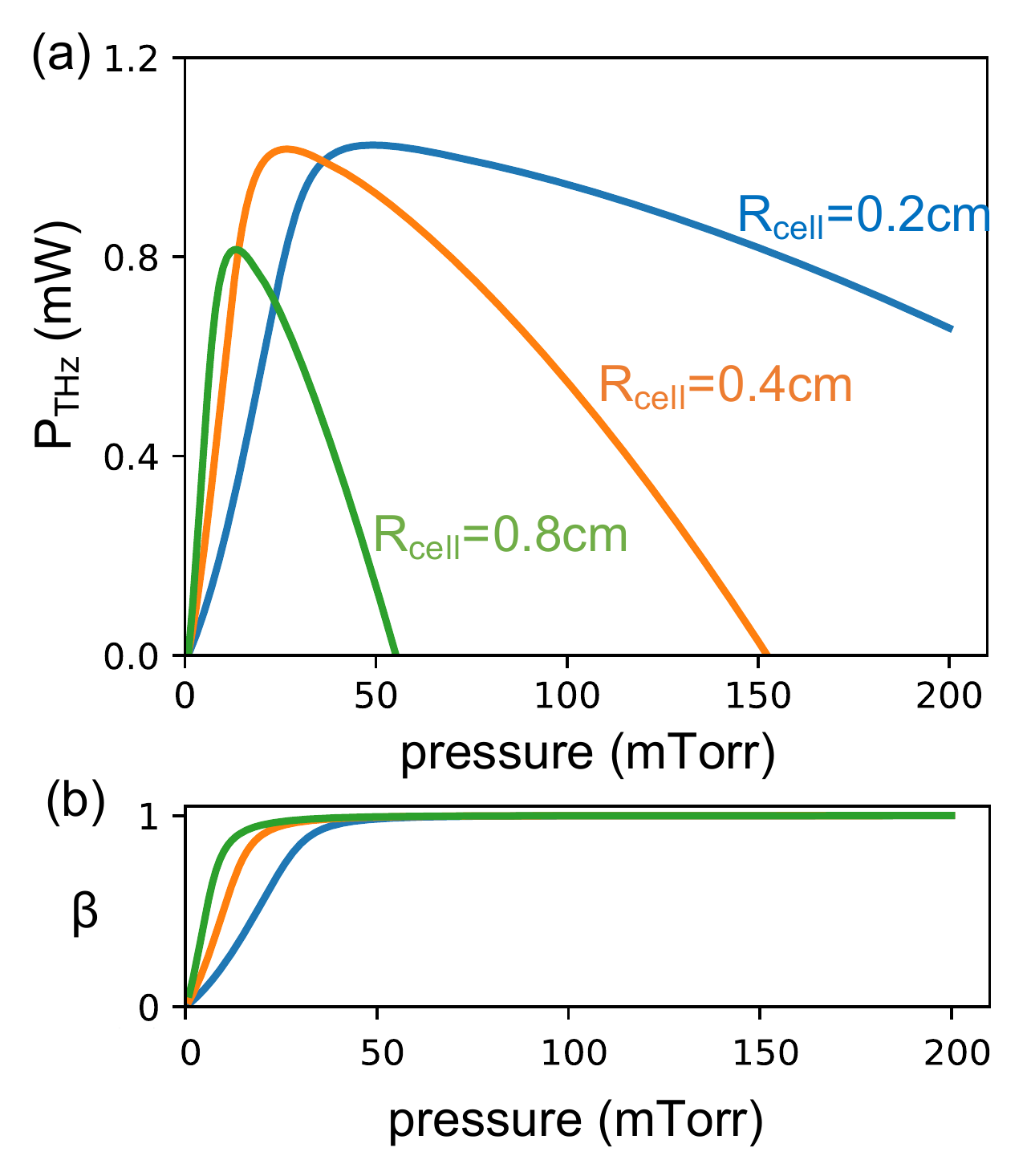}
    \caption{Terahertz output power and IR absorption $\beta$ as a function of pressure for cavity radii $R_\mathrm{cell}=$0.2, 0.4, and 0.8~cm with fixed cavity length $L = 15$~cm, front mirror transmission $T=0.016$, and 250~mW excitation of the $J_L=26$ transition.}
    \label{fig:PTHz_R}
\end{figure}

\Figref{fig:PTHz_R}(a) plots the corresponding QPML output powers $P_\mathrm{THz}$ as a function of pressure for the same three cavities using a 250~mW QCL pump. Both the peak pressure $p^\mathrm{opt}$ and maximum ``cutoff" pressure at which $P_\mathrm{THz}=0$ increase with decreasing cavity radius, a direct consequence of the quadratic dependence of $P_\mathrm{th}$ on $R_\mathrm{cell}$. As indicated above, the pressure $p^\mathrm{opt}$ at which peak power is achieved occurs as $\beta \to 1$. \Figref{fig:PTHz_R}(b) illustrates this by plotting $\beta$ over the same pressure range for all three radii.  Notice how $\beta$ increases faster with pressure for larger radii, a consequence of \eqref{eq:thzpower2} which shows that $P_\mathrm{THz}$ is proportional to $\beta$. Since the slope is larger in \figref{fig:PTHz_R}(a) at low pressures for cavities with larger $R_\mathrm{cell}$, cavities with smaller radii can operate over a much wider (and more forgiving) range of pressures than those with larger radii.

\begin{figure}[h]
    \centering
    \includegraphics[width=\columnwidth]{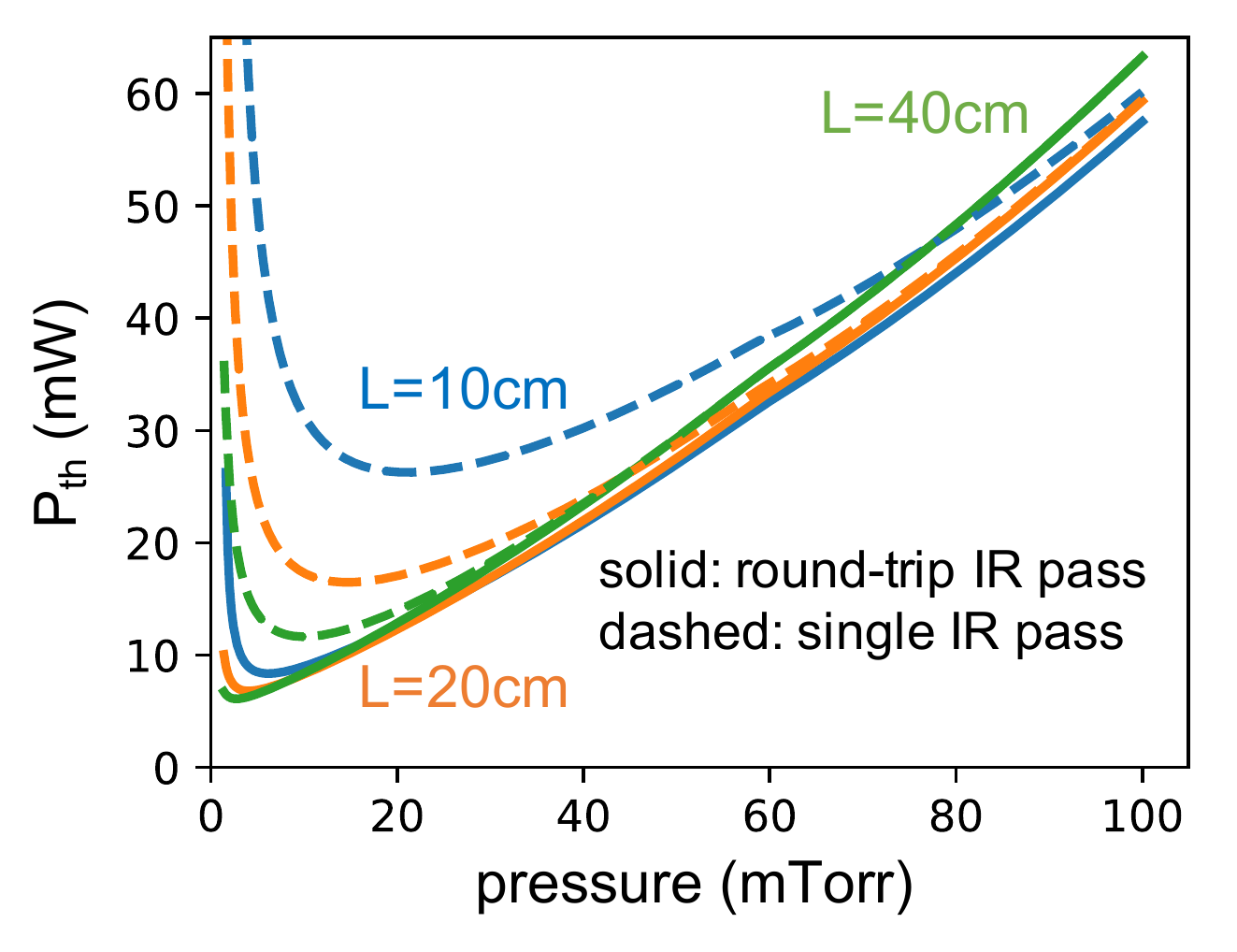}
    \caption{Solid lines: $P_\mathrm{th}$ versus pressure for cavity lengths $L=10$, 20, and 40~cm with fixed cavity radius $R_\mathrm{cell} = 0.25$~cm, front mirror transmission $T=0.016$, and pump excitation of the $J_L=26$ transition. Dashed lines: Same calculations but assuming only a single IR pass through the cavity without back reflection.}
    \label{fig:Pth_L}
\end{figure}

\textbf{Cavity length} affects the optimal operation pressure but doesn't greatly affect the threshold power in the regime considered here, $\alpha_o \ll \alpha_T \approx \alpha_\mathrm{cell}$, for which the limiting case of equation \eqref{eq:pth} becomes 
\begin{equation}
P_\mathrm{th}\approx (h\nu_\mathrm{IR}) (\pi R_\mathrm{cell}^2 T) (1.5k_\mathrm{DD}+k_w)/2\sigma\beta.
\end{equation}
The solid lines in \figref{fig:Pth_L} confirm this perhaps surprising result that $P_\mathrm{th}$ weakly depends on $L$. The reason is simple: the $\beta$ term captures the effects of multiple round trips and pump saturation, so the pump beam passes through the cavity as many times as necessary for all pump photons to be absorbed.  Indeed, the very slight dependence on cavity length comes from the additional losses of the IR pump in a short cavity as multiple round trips allow a small portion to be absorbed by the front and back metallic mirrors and leak through the output coupler ($T_\mathrm{IR}$) each round trip.  As noted above, this leakage ultimately limits the number of round trips possible in a given cavity, and this limit explains why $\beta <$ 1 at low pressures: too many pump photons are lost before being absorbed by the molecular gain medium. If the model only assumes a single pass for the IR pump, as shown in the dashed lines of \figref{fig:Pth_L}, $P_\mathrm{th}$ becomes larger and more sensitive to $L$ until the molecular pressure is high enough to absorb all the pump photons.

The fact that this metallic cavity geometry allows multiple round trips of the IR pump provides a significant advantage in reducing laser size. Most prior work on low-pressure OPFIR lasers used very long cavities to maximize absorption of pump photons in a single pass.  Regardless of geometry, notice that the peak power occurs as $\beta \to 1$, because this is the first pressure at which all the pump photons are absorbed by the molecular gain medium.  As pressure increases above $p^\mathrm{opt}$, pump photons are fully absorbed in the front portion of the cavity, leaving many molecules unpumped, an indication that the cavity is too long.  Operation at higher pressures was a principal motivation for the compact cavity concept~\cite{Everitt1986,Wang2018}. 

\begin{figure}[h]
    \centering
    \includegraphics[width=0.95\columnwidth]{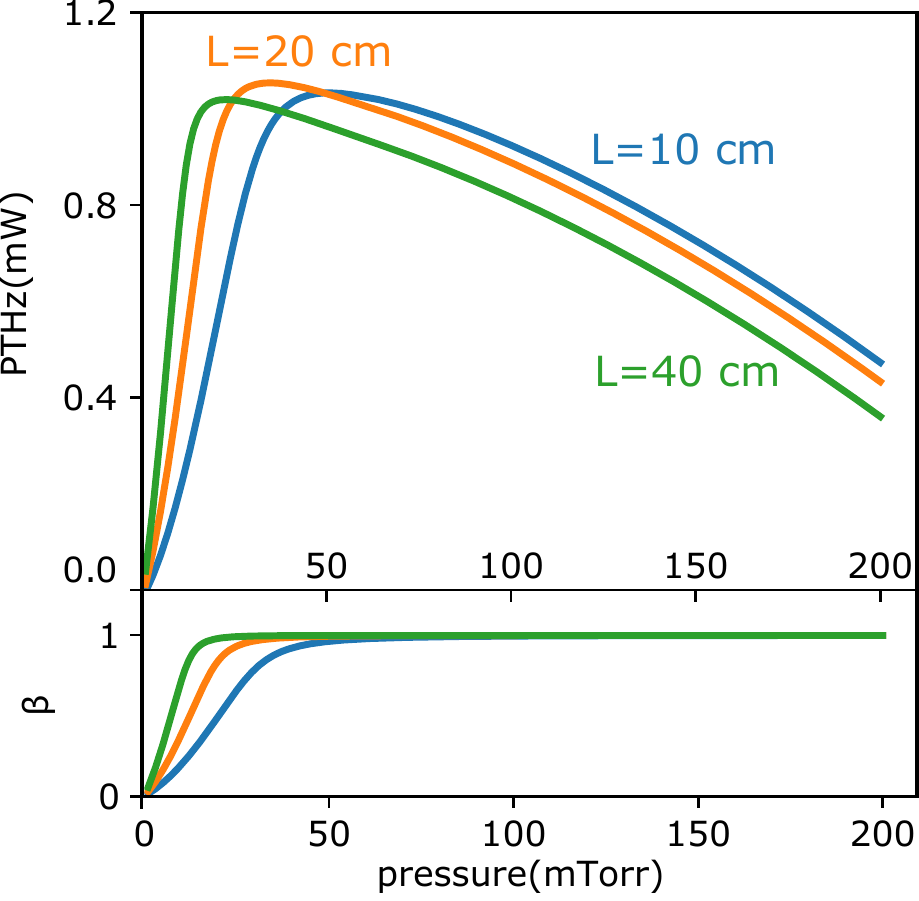}
    \caption{Terahertz output power and IR fractional absorption $\beta$ for cavity lengths $L=$10, 20, and 40~cm with fixed cavity radius $R_\mathrm{cell} =$ 0.25~cm, front mirror transmission $T=0.016$, and 250~mW excitation of the $J_L=26$ transition.}
    \label{fig:PTHz_L}
\end{figure}

\Figref{fig:PTHz_L}(a) plots output power as a function of pressure for cavity lengths of $L=10$, 20, and 40~cm and fixed radius $R_\mathrm{cell}=$~0.25~cm. These three curves are more similar than they are different, with comparable peak output powers occurring at the pressure where $\beta \to 1$, as can be seen in \figref{fig:PTHz_L}(b).  In short cavities the effects of pump saturation cause more round trips before complete absorption, so more of the IR pump is lost through leakage.  Consequently, shorter cavities require more pressure than longer cavities to reach their peak absorption and peak power. However, the high pressure behaviors and cutoffs are similar for all three cavity lengths, a consequence of how insensitive $P_\mathrm{th}$ is to $L$ (in contrast with its strong sensitivity to $R_\mathrm{cell}$ seen in \figref{fig:PTHz_R}).


\textbf{Output Coupler Transmission} $T$ at the emission frequency $\nu_\mathrm{THz}$ should be adjusted to maximize the output power $P_\mathrm{THz}$.  Note that $T_\mathrm{THz}$ is not the same as $T_\mathrm{IR}$, and how they differ depends on the type of output coupler used.  If it is a partially reflecting mirror, then the spectral dependence of the reflectivity and absorptivity of the window determines $T_\mathrm{THz}$ and $T_\mathrm{IR}$.  If it is a circular pinhole in an otherwise highly reflective circular mirror, then $T_\mathrm{IR}$ will likely be the ratio of the geometrical areas of the pinhole and mirror, while $T_\mathrm{THz}$ is approximated by the spatial overlap of the lasing mode with the pinhole. (The radiation through a very subwavelength hole would require an entirely different approach~\cite{bethe1944theory}.) Because the pinhole must be large enough for radiation to be emitted through ($\nu_\mathrm{cutoff} = 1.84 c / 2 \pi R_\mathrm{pinhole} \approx 88/R_\mathrm{pinhole}$~GHz for $R_\mathrm{pinhole}$ in mm), the requirement that $\nu_\mathrm{cutoff} < \nu_\mathrm{THz}$ determines the minimum possible value of $T$ for a given laser transition. To maintain a constant $T$ with increasing cavity radius, the pinhole diameter must also increase. Even larger pinhole diameters may be desired to reduce diffraction of the emitted terahertz beam, at the cost of larger values of $T$ and cavity loss, and lower values of $\beta$ due to more IR leakage.  

\begin{figure}[h]
\centering
    \includegraphics[width=\columnwidth]{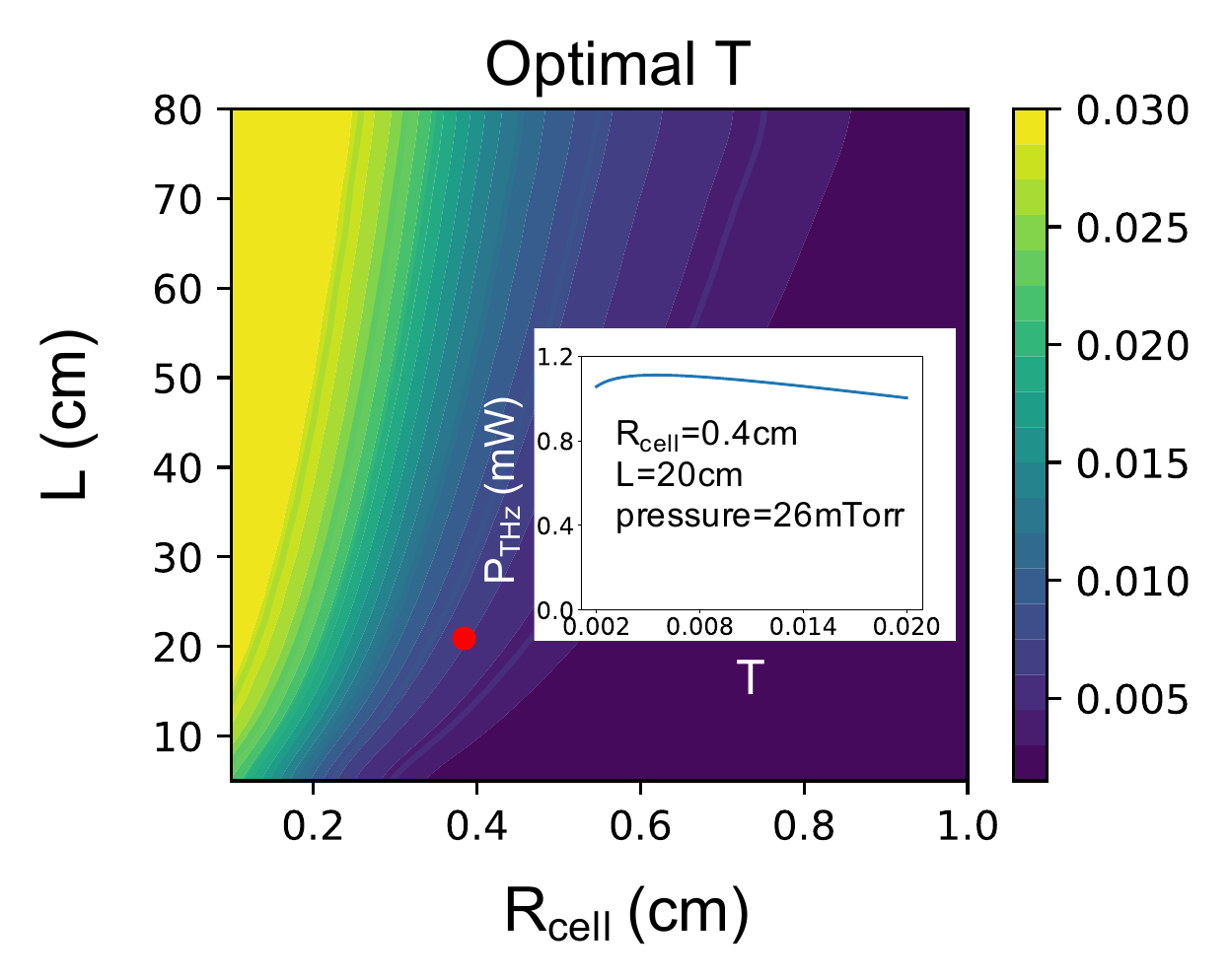}
    \caption{Optimal front mirror transmission coefficient $T$ as a function of  cavity radius $R_\mathrm{cell}$ and length $L$, each operating at $p^\mathrm{opt}$ for the $J_L=26$ transition pumped by a 250~mW QCL. The inset shows how the output power is relatively insensitive to $T$ for a cell with $R_\mathrm{cell}=0.4$~cm, $L=20$~cm, and  pressure$=26$~mTorr (red dot).}
    \label{fig:optT}
\end{figure}

Understanding these constraints, the optimal transmissivity may be obtained from $dP_\mathrm{THz} / dT = 0$, so using \eqref{eq:thzpower2} and \eqref{eq:pth}, we can derive
\eq{
T_\mathrm{opt} = \sqrt{\frac{4\beta \sigma \alpha_o L P_\mathrm{QCL}}{h\nu_\mathrm{IR} (\pi R_\mathrm{cell}^2) (1.5k_\mathrm{DD}+k_w)}} - 2\alpha_o L.
}
This complex equation contains most of the parameters already considered, suggesting that $T_\mathrm{opt}$ should be decided last, after optimal values for all these other parameters have been ascertained. As can be seen in \eqref{eq:thzpower2}, $P_\mathrm{THz}$ is rather insensitive to $T$ in the common regime for which $\alpha_\mathrm{cell} \approx \alpha_T$, fortuitously allowing us to pick a fixed front mirror transmissivity without sacrificing much laser efficiency. To illustrate this insensitivity, consider a cavity with $R_\mathrm{cell}=0.4$~cm, $L=20$~cm, pumped with a 250~mW QCL on the $J_L=26$ transition, and filled with 26~mTorr N$_2$O molecule, for which the optimal front mirror transmission coefficient is found to be $T_\mathrm{opt}=0.0056$. \Figref{fig:optT} explores how $T_\mathrm{opt}$ depends on $R_\mathrm{cell}$ and $L$ for this same transition for a range of cavity geometries, each operating at the optimal lasing pressure $p^\mathrm{opt}$. Notice very highly reflective output couplers are required, with $T < 1\%$ for all but the smallest diameter cavities. Interestingly, $T_\mathrm{opt}$ is rather insensitive to $L$ and decreases slowly with increasing $R_\mathrm{cell}$.  As the inset illustrates, the output power varies little as $T$ varies from 0.002 to 0.020, so a constant value such as $T=0.01$ may be set with little sacrifice in performance. 

\begin{figure}[h]
    \centering
    \includegraphics[width=\columnwidth]{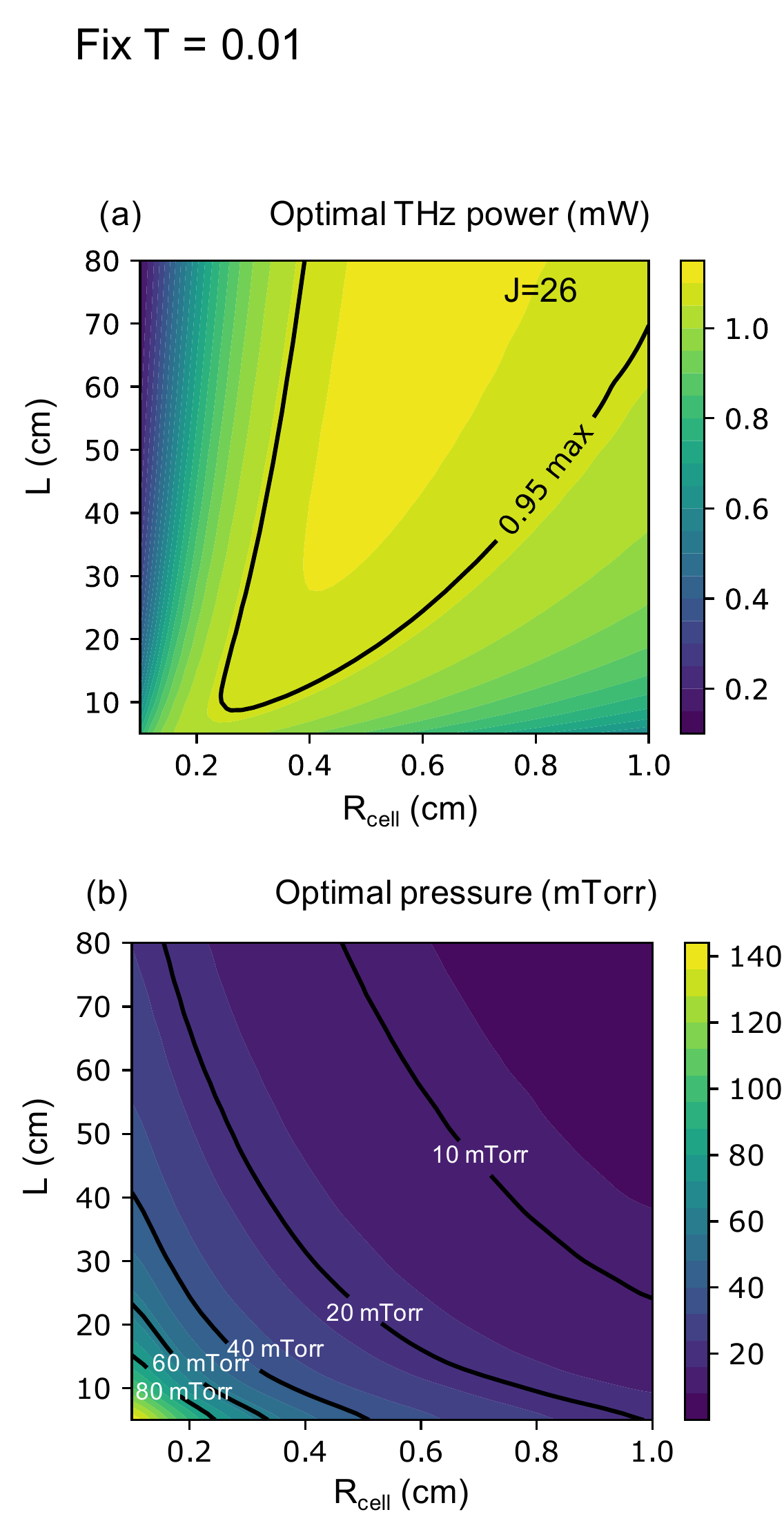}
    \caption{(a) Maximum terahertz power as a function of cavity radius and length for the $J_L=26$ transition with 250~mW QCL pump power. Each value of $P_\mathrm{THz}$ was obtained at $p^\mathrm{opt}$ with fixed $T=0.01$. (b) The corresponding optimal pressure $p^\mathrm{opt}$ for each cavity geometry in (a).}
    \label{fig:optcavity_J26}
\end{figure}

\textbf{Optimal cavity design} encompasses all these parameters to create conditions for maximum QPML power for a given QCL pump power.  The objective is to convert infrared photons to terahertz photons efficiently, thereby approaching the Manley--Rowe limit as closely as possible.  To illustrate how to ascertain the optimal cavity geometry using this model, \figref{fig:optcavity_J26}(a) plots the maximum output power as a function of cavity radius and length for the $J_L=26$ transition pumped by a 250~mW QCL with $T=0.01$. Half the Manley--Rowe limit \eqref{eq:PTHzShort} for this transition at $\nu_\mathrm{THz} = 0.672$~THz and $\nu_\mathrm{IR} = 67.3$~THz (2243.761 ~cm$^\mathrm{-1}$) is $P_\mathrm{THz} = 1.25$~mW. 

As can be seen in \figref{fig:optcavity_J26}(a), a wide range of cavity geometries approach this maximum power limit, indicating that QPMLs are rather tolerant of non-ideal cavity dimensions as long as they are operated at $p^\mathrm{opt}$. Since this optimal pressure for maximum laser power depends on cavity radius $R_\mathrm{cell}$ and length $L$, the powers $P_\mathrm{THz}$ plotted in \figref{fig:optcavity_J26}(a) were calculated for the corresponding $p^\mathrm{opt}$ plotted in \figref{fig:optcavity_J26}(b). As expected from the discussion above, $p^\mathrm{opt}$ decreases with increasing $R_\mathrm{cell}$ and $L$.  However, $P_\mathrm{THz}$ has a more complex dependence on radius and length, with maximum power achieved for an accommodating range of cavity radii ($0.4 < R_\mathrm{cell} < 0.8$~cm) and lengths ($L > 30$~cm) when operated at $p^\mathrm{opt}$.

In general, excellent performance may be obtained from compact cavity geometries ($R_\mathrm{cell} < 1$~cm, $L < 1$~m) operating at low pressures ($p^\mathrm{opt} < 100$~mTorr) with low front mirror transmission ($T \sim 0.01$). Although maximum power for this transition and pump power occurs in a cavity with $R_\mathrm{cell}=0.8$~cm, and $L=178$~cm, the 95\% contour line shows how forgiving the QPML is to non-ideal cavity geometries. This is of particular importance for practical applications where low volume geometries are favored. In this example, a compact laser cavity with $R_\mathrm{cell}=0.4$~cm and $L=20$~cm can achieve more than 95\% of the maximum output power but with $\sim 40\times$ smaller volume than the $R_\mathrm{cell}=0.8$~cm, $L=178$~cm ``optimal'' cavity.

\begin{figure}[h]
    \centering
    \includegraphics[width=\columnwidth]{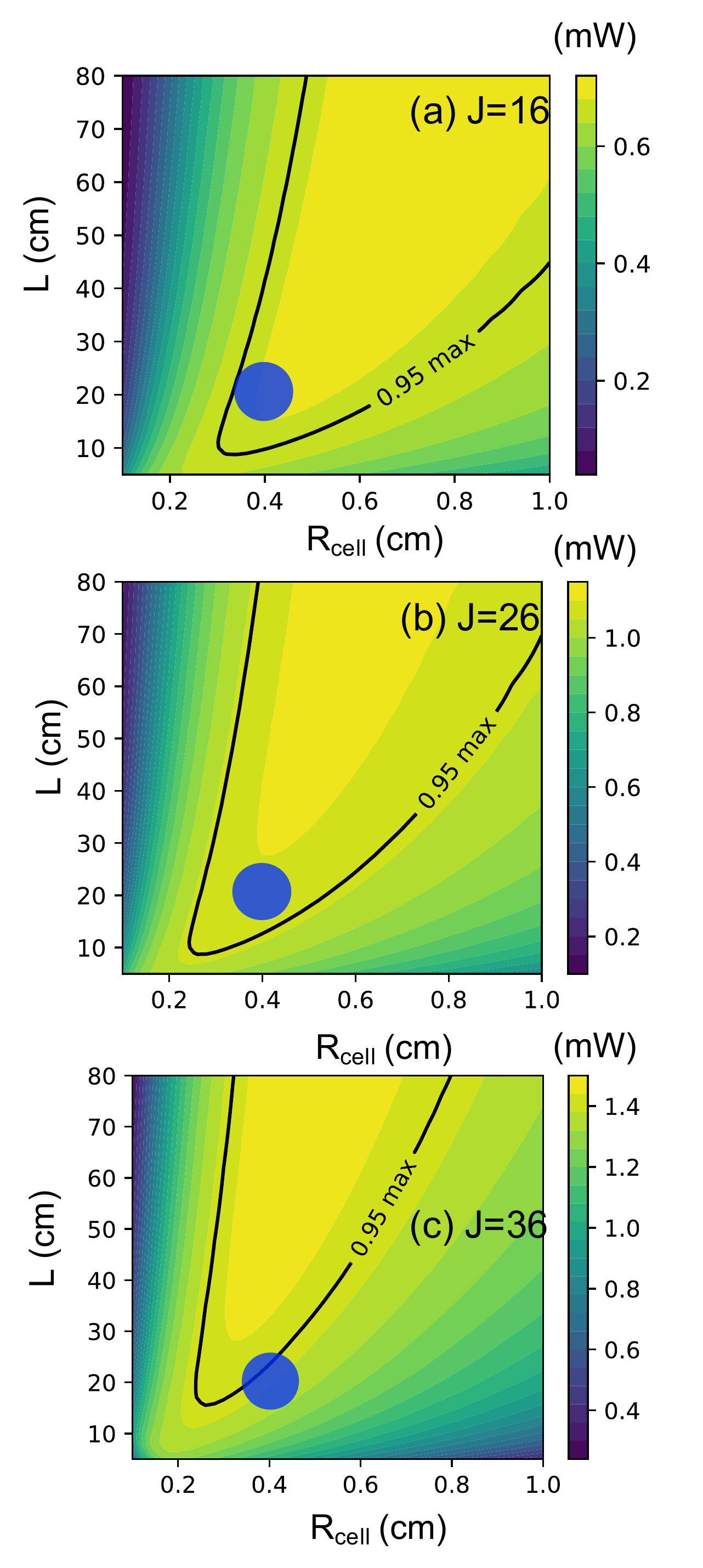}
    \caption{Maximum terahertz power for (a) $J_L=$ 16, (b) $J_L=$ 26, and (c) $J_L=$ 36 pumped by a 250~mW QCL. Each point is obtained by optimizing molecular pressure to maximize $P_\mathrm{THz}$ with fixed front mirror transmission $T=0.01$. The blue dot indicates a compact cavity geometry ($R_\mathrm{cell}=0.4$~cm, $L=20$~cm) that achieves at least 95\% of the maximum possible output power for all transitions.}
    \label{fig:optcavity_J16_26_36}
\end{figure}

Finally, let us consider how these parameters depend on the frequency of the QPML.  \Figref{fig:optcavity_J16_26_36}(a--c) plots the maximum terahertz power that can be achieved for the $J_L = $16, 26, and 36 transitions, with corresponding frequencies $\nu_\mathrm{THz} =$ 423, 672, and 921~GHz. The maximum output power increases with $J_L$ as expected by the Manley--Rowe effect~\cite{manley1956some}. Since $P_\mathrm{QCL} = 250$~mW $\gg P_\mathrm{th}$, the maximum possible powers may be estimated from \eqref{eq:PTHzShort} to be 0.79, 1.25, and 1.71 mW, respectively. Again, contour lines denote the broad range of cavity geometries that achieve 95\% of the maximum output power, and it can be seen that a single cavity geometry may be selected to achieve near-optimal performance for all three transitions and every transition between. Stated another way, the cavity geometry does not need change from transition to transition to achieve near-optimal performance, as long as the laser operates at $p^\mathrm{opt}$ for that transition. 

\rev{As compared to the optimal configuration for all transitions, one such compact geometry ($R_\mathrm{cell}=0.4$~cm, $L=20$~cm, $T = 0.01$), represented by the blue dot in \figref{fig:optcavity_J16_26_36}(a-c), only loses 5\% laser performance with a volume 10--100 times smaller.  The fact that a single compact, cigar-sized cavity can perform nearly optimally across the range of operational frequencies bodes well for the technological viability of the QPML concept.}

\rev{\section{Pump power dependence}}

\begin{figure}[h]
    \centering
    \includegraphics[width=\columnwidth]{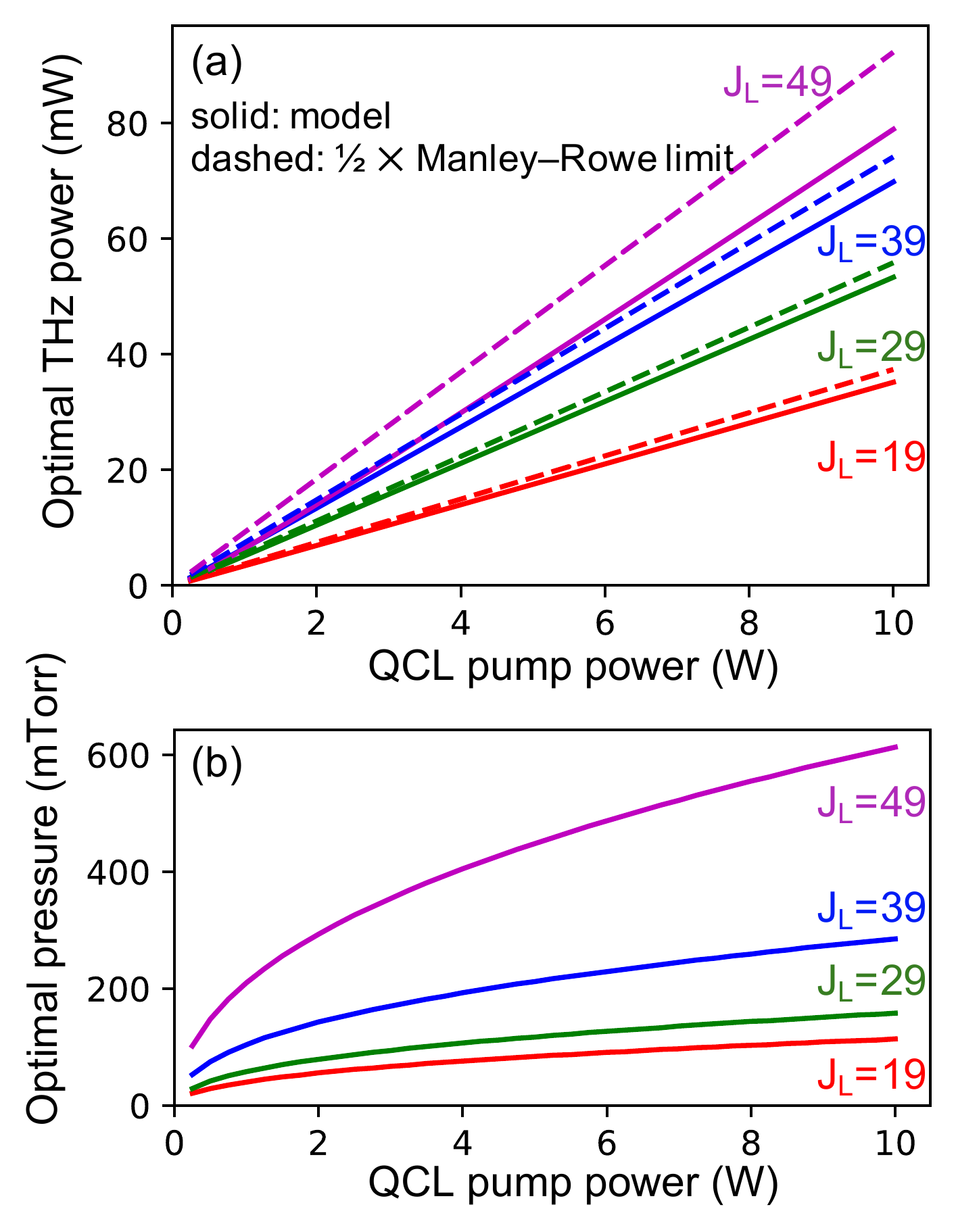}
    \rev{\caption{(a) Optimal THz power with varying QCL pump power for different transitions, and (b) the corresponding optimal molecular pressures, using the optimal cavity configuration ($R_\mathrm{cell}=0.4$~cm, $L=20$~cm, $T = 0.01$) obtained previously.}
    \label{fig:QCL10W_1}}
\end{figure}
\rev{Now that the optimal cavity geometry has been ascertained, we may revisit the one parameter held constant throughout this analysis:  QCL pump power.  Using the near-optimal compact cavity configuration $R_\mathrm{cell}=0.4$~cm, $L=20$~cm, $T = 0.01$, \figref{fig:QCL10W_1}(a) plots the maximum THz power (solid lines), and half the Manley--Rowe limit (dashed lines) as a function of QCL pump power for different transitions $J_L=$19, 29, 39, and 49. Because of the Manley--Rowe effect, the general trend is that optimal THz power increases with increasing $J_L$ and corresponding emission frequency. However, at the highest $J_L$ and frequency values, pump saturation and low infrared absorption increasingly cause the output power to deviate from the theoretical optimal (dashed lines), ultimately leading to the precipitous drop in power at the highest frequencies seen in \figref{fig:optmailPTHz_freq}.} 

\rev{Although the same cavity may be used for all these transitions, achieving optimal THz power as QCL pump power increases requires overcoming pump saturation by increasing the molecular pressure, as shown in \figref{fig:QCL10W_1}(b). For the same reason, the optimal pressure also increases with transition number $J_L$ to achieve sufficiently large $\alpha_\mathrm{IR}$ as the fractional population of $J_L$ drops.} 

\begin{figure}[h]
    \centering
    \includegraphics[width=\columnwidth]{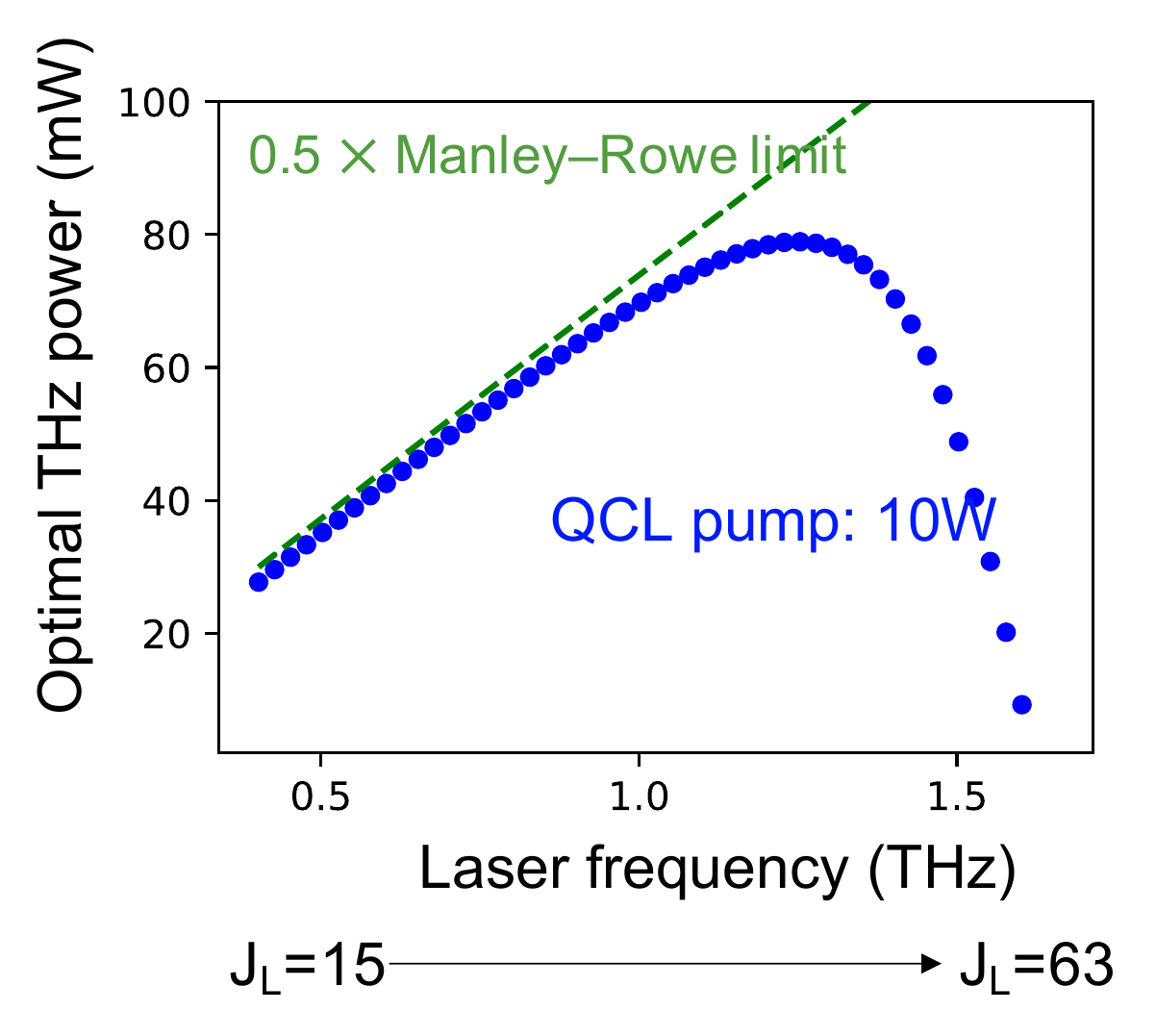}
    \rev{\caption{Optimal THz  power  for all transitions $J_L=15$ ($\nu_\mathrm{THz}=400$GHz) to $J_L=63$ ($\nu_\mathrm{THz}$=1.6THz) with a fixed 10W QCL pump power.}
    \label{fig:QCL10W_2}}
\end{figure}

\rev{With a fixed 10W QCL pump, \figref{fig:QCL10W_2} plots the optimal THz power for transitions $J_L=15$ ($\nu_\mathrm{THz}=400$GHz) to $J_L=63$ ($\nu_\mathrm{THz}$=1.6THz), similar to \figref{fig:optmailPTHz_freq} but with a much higher pump power. With a 10W QCL, the maximal THz power achievable is 79~mW for $J_L$ = 49 around 1.25~THz, far higher in frequency than the peak of the R-branch infrared absorption spectrum at $J_L$ = 15. Indeed, by comparing \figref{fig:optmailPTHz_freq} and \figref{fig:QCL10W_2} we see that as pump power increases, the increasing optimal pressure and the Manley--Rowe effect combine to increase the frequency at which peak THz power may be produced from a QPML. Multi-watt class QCLs are becoming commercially available, opening the exciting prospect that N$_2$O, NH$_3$, and other QPML gain media will soon be able to produce 10's, even 100's, of milliwatts in the region above 1 THz, a region currently devoid of comparably powerful, tunable, narrow linewidth sources~\cite{lewis2014review}.}

\section{Conclusion}
In this work, an improved three-level model of QCL-pumped molecular lasers is derived that accurately describes the nonlinear infrared absorption, including pump saturation, recycling of the pump beam within the cavity, and the essential collision physics. 
This new model, validated by the comprehensive model, allows us to study quantitatively how the laser performance is influenced by the pump strategies and cavity parameters. Our calculations find that a compact N$_2$O QPML cavity ($R_\mathrm{cell}=0.4$~cm and $L=20$~cm) is able to achieve 95\% of the maximum laser output power across almost all laser transitions with only 2.5\% the volume of a nominally optimized cavity. 

Throughout the paper N$_2$O is chosen as the gain molecule, but the model and the design  principles  derived  from  it  are  universal  and  easily  applied to other candidate linear molecules such as HCN, OCS, and CO. Extension of the model to very high pressures and more complex molecules such as CH$_3$F and NH$_3$ could also be made, requiring the inclusion of additional collisional relaxation processes, but the results derived herein may be applied to those molecules without loss of generality. \rev{As such, the improved three-level model introduced here may be used to assess and achieve the potential of powerful, broadly tunable QPMLs as a solution to the terahertz gap source problem.} 

\begin{acknowledgments}
This work was funded in part by the U.~S.~Army Research Office through the Institute for Soldier Nanotechnologies under award no.~W911NF-18-2-0048. We are also grateful to F.~Capasso, P.~Chevalier, and A.~Amirzhan for helpful conversations. 
\end{acknowledgments}





\bibliography{cavityopt}

\end{document}